\documentclass{ar-1col-S2O}
\usepackage[numbers]{natbib}
\usepackage{url}
\jname{The Annual Review of Biomedical Data Science.}
\jvol{9}
\jyear{2026}
\doi{10.1146/...}
\usepackage{amsmath}
\usepackage{amsthm}
\usepackage{amsfonts}
\usepackage{graphicx,psfrag,epsf}
\usepackage{enumerate}
\usepackage{times}
\usepackage{caption}
\usepackage{subcaption}
\usepackage{natbib}
\usepackage{multirow}
\usepackage{bm,algorithm} %!!check it see whether it can be used
\usepackage{url} % not crucial - just used below for the URL 
\usepackage{comment}

\usepackage{setspace}

\usepackage[pagewise]{lineno}
\usepackage{xr}

\makeatletter
\newcommand*{\addFileDependency}[1]{% argument=file name and extension
  \typeout{(#1)}
  \@addtofilelist{#1}
  \IfFileExists{#1}{}{\typeout{No file #1.}}
}
\makeatother
 
\newcommand*{\myexternaldocument}[1]{%
    \externaldocument{#1}%
    \addFileDependency{#1.tex}%
    \addFileDependency{#1.aux}%
}

\myexternaldocument{supplement}

\usepackage{epsf,pstricks,pst-node}
\usepackage{float}
\usepackage[outdir=./]{epstopdf}
\usepackage{verbatim} % comment

\newcommand{\bea}{\begin{eqnarray*}}
\newcommand{\eea}{\end{eqnarray*}}
\newcommand{\be}{\begin{eqnarray}}
\newcommand{\ee}{\end{eqnarray}}
\newcommand{\bay}{\begin{array}}
\newcommand{\eay}{\end{array}}
\newcommand{\bi}{\begin{itemize}}
\newcommand{\ei}{\end{itemize}}
\newcommand{\ben}{\begin{enumerate}}
\newcommand{\een}{\end{enumerate}}
\newcommand{\bcen}{\begin{center}}
\newcommand{\ecen}{\end{center}}

\usepackage[symbol]{footmisc}

\DeclareOldFontCommand{\bf}{\normalfont\bfseries}{\mathbf}

\allowdisplaybreaks

\usepackage{tabularx}

\begin{document}
\markboth{Wang et al.}
{Artificial Intelligence in
Image-based Cardiovascular
Disease Analysis}

% Title
\title{Artificial Intelligence in Image-based
Cardiovascular Disease Analysis}

\author{Xin Wang$^{1}$,  
Mingcheng Hu$^2$,
Connie W. Tsao$^3$, and
Hongtu Zhu$^2$
\affil{$^1$Department of Epidemiology and Biostatistics, College of Integrated Health Sciences and AI Plus Institute, the University at Albany, SUNY, NY, 12222; email: xwang56@albany.edu}
\affil{$^2$Department of Biostatistics  and Biomedical Research Imaging Center,  University of North Carolina, Chapel Hill, NC, 27514; email: htzhu@email.unc.edu}
\affil{$^3$Harvard Medical School, Beth Israel Deaconess Medical Center, Boston, MA, USA; (e-mail: ctsao1@bidmc.harvard.edu}
}

\begin{abstract}
Recent advancements in Artificial Intelligence (AI) have significantly influenced the field of Cardiovascular Disease (CVD) analysis, particularly in image-based diagnostics. 
Our paper presents an extensive review of AI applications in image-based CVD analysis, offering insights into its current state and future potential. 
We systematically categorize the literature based on the primary anatomical structures related to CVD, dividing them into non-vessel structures (such as ventricles and atria) and vessel structures (including the aorta and coronary arteries). 
This categorization provides a structured approach to explore various imaging modalities like Computed tomography (CT) and Magnetic Resonance Imaging (MRI), which are commonly used in CVD research. 
Our review encompasses these modalities, giving a broad perspective on the diverse imaging techniques integrated with AI for CVD analysis. %Additionally, we compile a list of publicly accessible cardiac image datasets and code repositories, intending to support research reproducibility and facilitate data and algorithm sharing within the community. 
We conclude with an examination of the challenges and limitations inherent in current AI-based CVD analysis methods and suggest directions for future research to overcome these hurdles.
\end{abstract}

\maketitle

%Table of Contents
% \tableofcontents
% \vspace{10pt}

%\spacingset{1.5} % DON'T change the spacing!
\section{Introduction}
\label{Introduction}

{C}{ardiovascular} disease (CVD) is the leading cause of mortality worldwide, accounting for an estimated 17.9 million deaths annually, as per the World Health Organization (WHO). 
CVD, encompassing heart disease and stroke, is the primary cause of death in the United States. Heart disease alone was responsible for approximately 697,000 deaths in 2020, according to the Centers for Disease Control and Prevention (CDC). 
These statistics underscore the profound impact of CVDs on global public health and underscore the critical need for enhanced medical research and healthcare strategies in this domain. Consequently, timely and accurate diagnosis, risk assessment, and treatment planning are essential to mitigate their impact and improve patient outcomes \cite{update2017heart,sun2023social}.

% %--------------------
%  \begin{figure}[t] % use t for all the figs and tabs
% \centering
% \includegraphics[width=0.95\linewidth]{fig/aicvdintro.png}
% \caption{\it This figure illustrates our comprehensive approach to exploring CVDs. We categorize these diseases according to the primary anatomical structures they impact and their respective functions. Our analysis also encompasses a variety of medical imaging modalities employed in the diagnosis and study of these diseases. Furthermore, we highlight the significant role of Artificial Intelligence (AI) in enhancing image-based cardiovascular disease analysis, showcasing the integration of advanced AI techniques with traditional imaging methods. 
% }
% \label{motivation}
% \end{figure} 
% %--------------------

Cardiac imaging encompasses a wide range of modalities used to visualize and evaluate the heart’s structure and function \cite{li2023multi}. These methods provide essential information about   heart's chambers, blood vessels, and associated tissues, supporting several key roles in clinical care and research. 
(i) Cardiac imaging aids in identifying biomarkers linked to the risk, progression, and treatment response of heart diseases \cite{vasan2006biomarkers}. Indicators such as carotid artery wall thickness can act as a biomarker for atherosclerosis and predict future cardiovascular events. It can also reveal changes in heart tissue or blood flow that serve as biomarkers for various heart conditions \cite{michelhaugh2023using}, enhancing disease understanding and guiding treatment decisions.
(ii) Cardiac imaging is integral to both the planning and monitoring of treatments for various cardiovascular diseases, including coronary artery disease, heart failure, and valvular heart disease. Techniques like angiography are instrumental in visualizing blood flow in coronary arteries and detecting obstructions, while echocardiography evaluates heart function and detects structural or motion abnormalities. These imaging methods enable physicians to devise optimal treatment strategies and adjust them over time based on patient progress.
(iii) In the realm of clinical trials, cardiac imaging plays a pivotal role in evaluating the safety and efficacy of new medical therapies. It allows for real-time visualization of the heart's internal structures and functions and can track morphological or functional changes induced by new drugs or treatments. This capability is essential for determining the effectiveness and safety of treatments before their widespread adoption.

The incorporation of Artificial Intelligence (AI) in medical imaging has significantly transformed the analysis of CVDs \cite{bizopoulos2018deep,krittanawong2023deep}. 
AI, particularly deep learning methods \cite{lecun2015deep,mascarenhas2021comparison,zhou2021review}, has markedly improved the precision, efficiency, and objectivity of interpreting cardiovascular images \cite{shen2017deep,yang2023application}. The integration of AI with various cardiovascular imaging techniques, including MRI, CT, X-ray, and ultrasound, has enabled more comprehensive and dynamic evaluations of cardiovascular structures and functions \cite{he2023blinded,litjens2019state}. Recent AI advancements have facilitated breakthroughs in cardiovascular imaging tasks, including segmentation, disease classification, risk prediction, and clinical decision-making support for treatment planning \cite{shen2017deep,shin2016deep}. These developments highlight AI's crucial role in enhancing the fight against cardiovascular diseases.

%The benefits of AI in image-based cardiovascular disease analysis are manifold. 
%Firstly, AI algorithms can enhance the accuracy and efficiency of image interpretation, reducing inter-observer variability and facilitating early disease detection. Secondly, AI can aid in risk stratification and prognosis prediction, providing valuable insights for personalized treatment planning and patient management. 
%Thirdly, AI algorithms can improve workflow efficiency by automating time-consuming tasks, allowing clinicians to focus on complex cases and critical decision-making. 

% in this work, structure-function-disease

\begin{figure*}[t] % use t for all the figs and tabs
\centering
\includegraphics[width=0.98\linewidth]{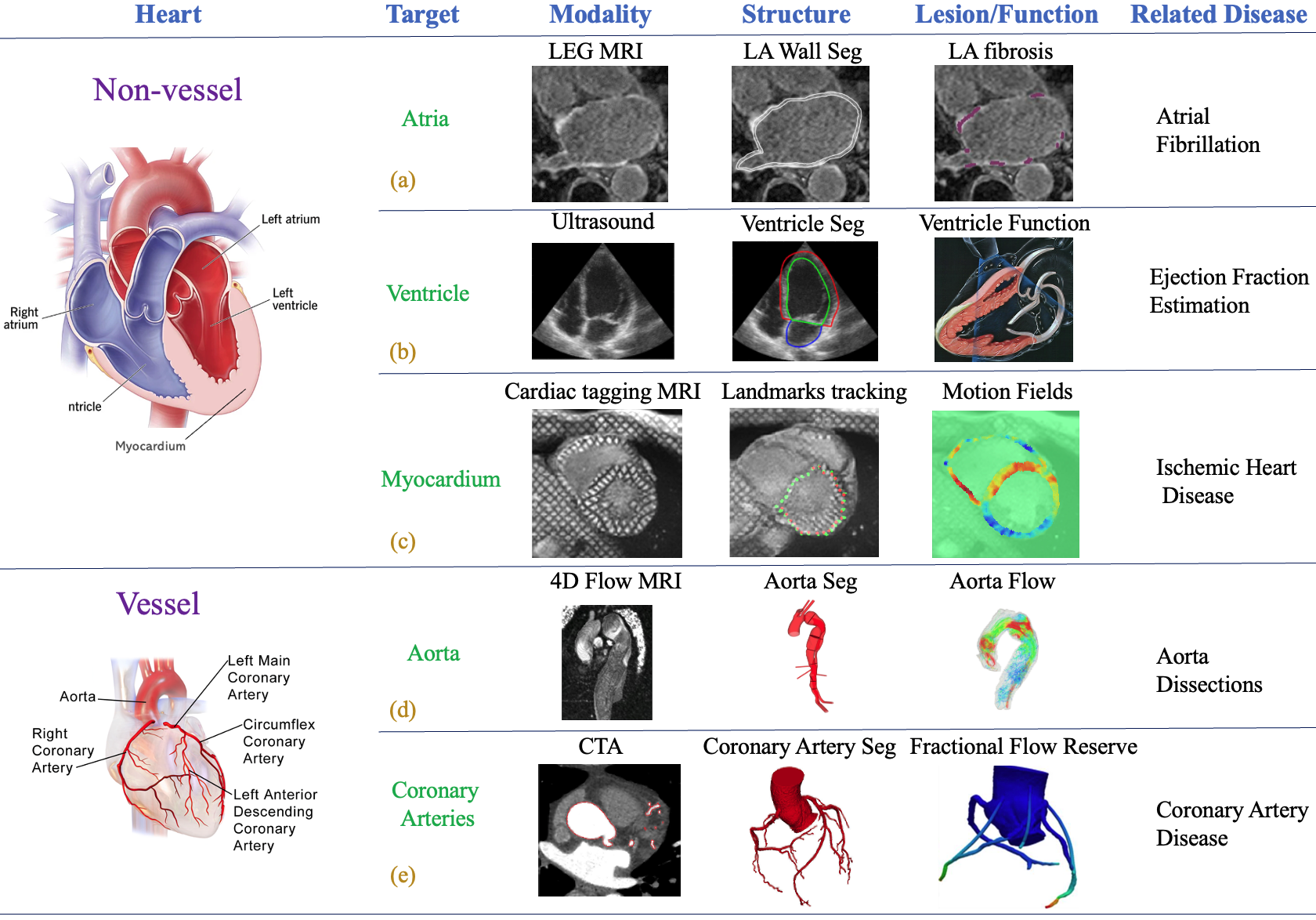}
\caption{\it This overview figure illustrates our comprehensive approach to exploring cardiovascular diseases (CVDs). We categorize these diseases according to the primary anatomical structures they impact and their respective functions. Our analysis also encompasses a variety of medical imaging modalities employed in the diagnosis and study of these diseases.
The example categorizes CVDs by highlighting the key anatomical structures involved and their functions.
\textbf{Top:} It includes examples of non-vessel anatomical structures, such as (a) atria \cite{li2022medical}, (b)  ventricles  \cite{leclerc2019deep} and (c) Myocardium \cite{ye2021deeptag}.
\textbf{Bottom:} It presents vessel structures, including the (d) aorta  \cite{pirola20194} and (e) coronary arteries  \cite{serrano2023coronary}, \cite{kolli2018image}. 
}
\label{overview}
\end{figure*}

In this comprehensive review, we delve into the recent advancements and trends in Artificial Intelligence (AI) applied to the image-based analysis of cardiovascular diseases (CVDs), as illustrated in Figure \ref{overview}. Our approach is threefold:
{\it (i)  Categorization Based on Anatomical Structures}: We systematically organize an extensive collection of CVD research according to the primary anatomical structures impacted and their functions. This categorization creates two principal groups: non-vessel structures such as atria and ventricles, and vessel structures, including the aorta and coronary arteries (See Figure \ref{overview}). This classification is based on similarities in both the anatomical features and the analysis techniques employed. For example, vessel structure analyses often involve methods like vessel tracing and stenosis degree estimation.
{\it (ii) Integration with Other Data Types}: Beyond traditional medical imaging, our review includes studies that integrate imaging with additional data types, like genomics, for a more holistic analysis of CVDs \cite{zhao2023heart}. This expansive coverage, which includes a variety of cardiac imaging structural analysis tasks—such as chamber segmentation, coronary artery segmentation, and coronary artery branch labeling—sets our review apart. Additionally, we explore functional simulation methods, such as fractional flow reserve (FFR), a vital diagnostic measure in cardiology for assessing the severity of coronary artery stenosis.
{\it (iii) 
Large-Scale Population-Based Studies}: The review expands to encompass large-scale studies using AI for image-based analysis of CVDs. The merger of imaging genetics with AI in cardiovascular disease research represents a significant leap forward. It provides researchers with the tools to decode the genetic foundations of these diseases, significantly improving diagnosis, treatment strategies, and prognostic evaluations \cite{zhao2023heart}.

%By combining imaging genetic information with advanced AI algorithms, we can enhance our understanding of disease mechanisms, develop personalized treatment strategies, and ultimately improve patient outcomes in cardiovascular medicine.

% Existing survey and drawbacks

There exists a range of survey papers that have provided overviews of AI applications in cardiovascular imaging tasks, such as segmentation \cite{chen2020deep,fu2021review}, or that have focused on specific diseases \cite{li2022medical}, types of CVD like congenital heart disease (CHD) \cite{jone2022artificial}, or singular medical image modalities \cite{jafari2023automated}. However, these surveys have typically lacked a comprehensive approach that encompasses both structural and functional aspects of cardiovascular analysis. Notably, critical techniques like coronary flow analysis \cite{berry2014fractional} are often omitted. 
In our survey, we address this gap by discussing cardiac imaging modalities along with their unique characteristics, advantages, and the diseases they are most relevant to. 
Our approach covers both the anatomical aspects (such as chamber volume and coronary artery morphology) and functional aspects (including cardiac motion and blood flow) of cardiovascular imaging. 
%Additionally, we provide a summary of publicly available datasets and code repositories to enhance research reproducibility, foster collaboration, and accelerate scientific advancements in this field.
Furthermore, our survey engages in insightful discussions about the current challenges in AI-based cardiovascular imaging and outlines potential directions for future research in this rapidly evolving domain.
%---------------
% TODO: Delete space for Ischemic Heart Disease in figure
%---------------
The key contributions of our survey are outlined as follows:

\begin{itemize}
\item We provide the first comprehensive survey that thoroughly examines the use of AI in cardiac imaging. This survey uniquely emphasizes both the structural and functional aspects of the heart, encompassing a broad range of cardiac diseases. Our approach offers a holistic view of the transformative role of AI in cardiac imaging, providing valuable insights into its capabilities and limitations, thereby setting the stage for future innovations in this field.  
% \item A significant aspect of our work is the integration of cardiac imaging with large-scale, population-based studies through AI technologies. We explore how AI can effectively analyze image-based CVD data in combination with genetic, lifestyle, and environmental health factors. This multidisciplinary approach results in a nuanced understanding of CVD, considering the interplay between medical imaging data and broader genetic and environmental factors. Such comprehensive analysis is crucial for developing more personalized and effective strategies for CVD prevention and management.
% \item Our survey also sheds light on potential future research directions, particularly emphasizing the need for integrating diverse data types. This includes combining environmental and genetic data to develop sophisticated AI models for addressing complex clinical challenges in heart disease. By leveraging such multi-scale data, AI models can significantly improve our understanding of heart diseases, aiding in early detection, accurate diagnosis, and the formulation of effective treatment plans. This suggests that future advancements in AI will likely concentrate on leveraging these diverse data modalities to address intricate clinical questions in cardiology, aiming to enhance patient outcomes significantly. 

\item Another key contribution of our work lies at the intersection of cardiac imaging and population studies through AI, where we systematically integrate image-based CVD data with multi-scale determinants of health, including genetic predispositions, lifestyle factors, etc. This multidisciplinary paradigm enables a holistic understanding of CVD pathogenesis by deciphering the interactions between imaging phenotypes and broader biological/social contexts. 
Such integration is not only critical for personalizing prevention strategies but also reveals unmet needs in current research: future advances must prioritize developing AI architectures capable of synthesizing these heterogeneous data modalities to address clinically complex scenarios.
\end{itemize}

% Structure of this review +++++++++
% The remaining content of this paper is organized as follows. After introducing cardiac imaging modalities as the background in Section~\ref{CardiacImaging}, 
% we describe the heart structures (non-vessel and vessel), the common diseases, and the related AI works in Section~\ref{heartmethod}. 
% We also review the population-based cardiovascular imaging analysis works in Section \ref{PopulationIG}.
% In Sections \ref{datacode}, we summarize the public datasets and code. 
% We discuss the limitations of the current AI-based methods and the future works in Section \ref{futuredirection}. 
% Finally, we conclude the paper in Section~\ref{conclusion}.

%\xin{think about what's the major innovation feature of this survey. }

\section{Background - Cardiac Imaging}
\label{CardiacImaging}

%\xin{structure, function for what disease, the relation between image and diseases. }

%\xin{clinical trial, treatment, bio-marker, research}

In this section, we delve into the commonly employed modalities in cardiac imaging. Depending on the specific clinical requirements and the information needed by healthcare professionals, these imaging techniques are either used in combination or selectively chosen to provide the most accurate and comprehensive assessment of cardiac health.

% helps guide treatment decisions, monitor disease progression, and assess the effectiveness of interventions, such as coronary artery disease, heart failure, congenital heart defects, valvular heart disease, and arrhythmias. 

%Research: Heart medical imaging plays a critical role in cardiovascular research. 
%Researchers use medical imaging to study the heart's anatomy, function, and blood flow in detail. 
%This can lead to a better understanding of how the heart works, how diseases develop, and how they can be prevented or treated. 
%Researchers might use heart medical imaging to study the effects of new drugs or interventions on heart function or to explore the genetic and environmental factors that contribute to heart disease. This research can lead to new insights and potentially pave the way for new treatments and therapies.

% Overall, heart medical imaging is an indispensable tool in modern medicine and research. It provides a non-invasive, detailed view of the heart that can help in the diagnosis, treatment, and management of heart-related conditions. It also plays a crucial role in advancing our understanding of the heart and developing new treatments for heart disease.

%\xin{a fig, show examples for each modaility}

% segmentation Fractional Flow Reserve (FFR) 

\subsection{Electrocardiogram (ECG)}
An electrocardiogram (ECG) \cite{shenoy2024novel,christensen2024vision} is a critical medical test that captures the heart's electrical activity \cite{berkaya2018survey}. 
It graphically represents this activity over time, with the x-axis denoting time and the y-axis the amplitude of the electrical signal. The ECG can be perceived as a form of medical imaging, mapping the heart's electrical impulses through wave patterns on a graph \cite{li2020survey}.  
To obtain an ECG, electrodes are strategically placed on the patient's skin to record electrical impulses generated during the heart's contraction and relaxation cycles. Each heartbeat begins with an electrical impulse originating from the heart's sinoatrial (SA) node. This impulse travels through the heart muscle, prompting it to contract and pump blood. The ECG machine detects and amplifies these impulses, transforming them into waveforms that offer insights into heart function.
%
% ECG is an indispensable diagnostic tool in cardiovascular medicine, offering clinicians immediate insight into the heart's electrical function \cite{darmawahyuni2022deep}. It is essential for diagnosing a variety of heart-related conditions, including:
% (i)  
% Heart Rhythm Disorders: An ECG is effective in identifying abnormal heart rhythms or arrhythmias, such as atrial fibrillation or ventricular tachycardia.
% (ii) Heart Attacks: The ECG is instrumental in recognizing patterns indicative of myocardial ischemia (reduced blood flow) or infarction (tissue death due to lack of oxygen), which are key markers of heart attacks.
% (iii) Structural Heart Abnormalities: An ECG can reveal structural heart changes that suggest conditions like cardiomyopathy or hypertrophy, where the heart muscle is enlarged or altered \cite{ryu2023coat}. 

%---------------------------------
\subsection{Cardiac X-ray}

Cardiac X-ray imaging (can be static images and/or real-time fluoroscopic imaging), often employed in coronary angiography, is a diagnostic technique that uses X-rays to produce two-dimensional images of the heart and blood vessels \cite{matsuoka2022deep}. 
This method is particularly useful in emergency settings for initial assessments due to its quick execution and relatively low radiation exposure, making it safer for a wide range of patients \cite{ccimen2016reconstruction}. 
It is excellent for identifying coronary artery disease (CAD) by directly visualizing the coronary arteries and informing treatment strategies like angioplasty or stenting to restore heart muscle blood flow \cite{yu2018coronary}.

\subsubsection{Digital Subtraction Angiography (DSA)} This advanced technique in cardiac X-ray imaging enhances image clarity by digitally removing background structures, focusing solely on blood vessels. The use of contrast dye in conjunction with rapid image acquisition results in high-resolution images that are crucial for identifying vascular abnormalities \cite{crabb2023deep}.

\subsubsection{Interventional Coronary Angiography (ICA)} ICA is a specialized application of Cardiac X-ray imaging used for interventional procedures such as percutaneous coronary intervention (PCI). During PCI, a catheter is inserted to treat blockages in coronary arteries, often employing techniques like angioplasty and stenting. The X-ray system plays a crucial role in guiding the catheter's placement and monitoring the progress of the intervention \cite{niimi2022machine}.

\subsection{Cardiac Computed tomography (CT) }

CT, utilizing advanced X-ray technology, provides detailed cross-sectional, three-dimensional images of the heart and surrounding vessels. Compared to Cardiac X-ray imaging, Cardiac CT involves higher levels of radiation exposure but offers more detailed and comprehensive views. This imaging method is particularly valuable for assessing cardiac anatomy, with a primary focus on the coronary arteries. It plays a crucial role in identifying blockages or narrowing within these arteries, which are key indicators of coronary artery disease \cite{garg2023role}. In the following sections, we will explore various Cardiac CT modalities and their specific applications in cardiac imaging \cite{gu2021cyclegan}.

\subsubsection{Coronary CT Angiography (CTA)}
Non-contrast CT imaging utilizes tissue density differences to create images, allowing distinction between soft tissues, calcium, fat, and air. This is useful for estimating calcium presence in coronary arteries. In contrast-enhanced coronary CTA, acquired post-contrast agent injection, the imaging provides detailed views of cardiac chambers, vessels, and coronaries, effectively detecting non-calcified coronary plaques \cite{lin2022deep}. It's widely employed for diagnosing coronary artery disease (CAD), coronary anomalies, and pre-surgical evaluations for coronary bypass. Additionally, it assesses stent patency post-implantation \cite{xu2023coronary}.

\subsubsection{Calcium-Scoring Heart Scan}
This scan, also known as coronary artery calcium (CAC) scoring, is a specialized X-ray test that measures calcifications in coronary arteries, indicative of coronary artery disease (CAD). These calcifications, even before symptom onset, signal potential heart-related events. The derived calcium score helps in risk assessment for CAD, with higher scores suggesting increased risk of heart attacks \cite{eng2021automated}.

\subsubsection{Functional Cardiovascular CT}
This CT form is essential for non-invasive evaluation of heart and vascular structures. Known for producing detailed body cross-sections, it offers insights into both structural and functional aspects of the cardiovascular system. Functional Cardiovascular CT is instrumental in assessing myocardial perfusion and ventricular function, useful in evaluating ischemic heart disease, cardiomyopathies, and heart failure \cite{peper2020functional}.

\subsubsection{Cardiac Dual-Energy CT (DECT)}
DECT applies dual-energy imaging principles to the heart and vasculature, utilizing two X-ray energy spectra to yield detailed information on cardiac tissues and pathologies. It enhances plaque characterization in arteries and differentiates tissue types, aiding in myocardial infarction diagnosis and tumor identification \cite{bruns2020deep}.

\subsection{Cardiac Ultrasound (Echocardiography)}

Cardiac ultrasound imaging is a fundamental tool in cardiovascular assessment, employing sound waves to produce real-time images of the heart \cite{zhou2021artificial, liu2023deep}. These images reveal critical details about the heart's size, shape, function, and blood flow patterns \cite{aly2021cardiac}. 
Echocardiography's non-invasive approach, combined with the absence of ionizing radiation and its capability for immediate imaging, makes it a preferred choice for initial diagnosis in a range of cardiac conditions \cite{lu2023ultrafast}. 
It is crucial in identifying and managing various heart diseases, such as valvular heart disease, cardiomyopathies, and congenital heart disease \cite{ghorbani2020deep,jone2022artificial}.
For the vessels, Intravascular Ultrasound (IVUS) is a medical imaging methodology used in cardiology to visualize the inside of the heart's coronary arteries from within the artery itself \cite{xu2020fundamentals}. 
%It involves using a specially designed catheter with a miniaturized ultrasound probe attached to its distal end. 
The catheter is threaded through the coronary vasculature to the area of interest, and ultrasound is used to produce detailed images of the coronary arteries' interior walls \cite{de2002intravascular}.

\subsection{Nuclear Cardiology}
%\noindent\textbf{Myocardial Perfusion Imaging (MPI):} MPI is pivotal for assessing blood flow to the heart muscle, particularly useful in detecting regions that receive insufficient blood due to coronary artery blockages. This technique provides critical insights for diagnosing and managing coronary artery disease \cite{alskaf2022deep}.
 
Nuclear imaging in cardiology utilizes small amounts of radioactive tracers injected into the patient's bloodstream. These tracers emit gamma rays, captured by specialized cameras to generate images of the heart's structure and function \cite{tamaki2024current}. 
Nuclear cardiology, including single-photon emission computed tomography (SPECT) and positron emission tomography (PET), have been a key non-invasive imaging modality for patients with known or suspected cardiovascular disease.

\subsubsection{Cardiac Single-Photon Emission Computed Tomography (SPECT)} 
SPECT is a nuclear imaging technique that provides three-dimensional images of blood flow to organs and tissues, widely used in the assessment of heart diseases. In cardiac applications, SPECT helps evaluate myocardial perfusion—how well blood flows through the heart muscle—by allowing clinicians to see areas with reduced blood flow, which may indicate CAD or previous heart damage \cite{apostolopoulos2023deep}.

\subsubsection{Positron Emission Tomography (PET)} PET is a highly advanced and non-invasive imaging technique used in the evaluation of heart diseases. It utilizes a small amount of radioactive material, a PET scanner, and a computer to evaluate the function and metabolism of the heart. This technique stands out for its exceptional accuracy in detecting coronary artery disease, assessing myocardial perfusion (blood flow to the heart muscle), and evaluating heart function. Cardiac PET is particularly effective in identifying areas of reduced blood flow, differentiating between viable and non-viable heart muscle, and diagnosing various cardiac conditions. Its high sensitivity and specificity make it a valuable tool in the management of heart diseases \cite{slomka2021quantitative}.

\subsection{Cardiac Magnetic Resonance Imaging (MRI)}

Cardiac MRI is a non-invasive imaging technique that uses a powerful magnetic field and radio waves to produce detailed images of the heart, offering crucial insights into its structure, function, and blood flow \cite{saeed2015cardiac,bello2019deep,wang2024screening}. We discuss several MRI modalities commonly used in cardiac imaging.

\subsubsection{Cine MRI} Cine MRI captures a series of images across the cardiac cycle to visualize the beating heart in real time, allowing evaluation of heart structure, ventricular size, wall motion, and ejection fraction. It's essential in detecting conditions like ventricular hypertrophy, myocardial infarction, and various valvular disorders \cite{kustner2020cinenet, campello2021multi}.

\subsubsection{Cardiac tagging magnetic resonance imaging (t-MRI)} 
Cardiac tagging magnetic resonance imaging (t-MRI), also known as myocardial tagging or t-MRI, offers a unique approach to evaluating myocardial movement and deformation. During imaging, a grid-like pattern or "tags" are superimposed on the heart muscle, revealing intricate details about myocardial function. These tags change shape as the heart beats, providing valuable insights into both global and regional cardiac function. Regarded as the gold standard for assessing regional myocardial deformation and strain, cardiac tagging is pivotal in the diagnosis, management, and research of various heart conditions, particularly in diseases such as Ischemic Heart Disease and Dilated Cardiomyopathy \cite{ye2021deeptag}.

\subsubsection{Late Gadolinium Enhancement (LGE) MRI}
 LGE MRI is a crucial imaging technique that enhances the visibility of myocardial damage or fibrosis by using a gadolinium-based contrast agent. It plays a key role in evaluating myocardial viability and identifying pathological tissues, such as scarred areas. This method is particularly effective in diagnosing and assessing various cardiac conditions, including myocardial infarction, myocarditis, and different types of cardiomyopathies. LGE MRI is also increasingly recognized as a valuable tool for assessing scar tissues in patients with atrial fibrillation (AF), offering an advanced alternative for detailed cardiac evaluation \cite{li2022medical}.

\subsubsection{T2-weighted MRI Images}
T2-weighted MRI stands out for its ability to provide enhanced contrast by distinguishing tissues based on water content and physiological state. This feature is particularly valuable in identifying myocardial abnormalities, such as inflammation, edema, or ischemia. T2-weighted MRI is especially effective in detecting myocardial edema caused by inflammation or acute ischemia, offering critical insights for diagnosis and treatment \cite{ren2022comparison}.

\subsubsection{Myocardial Perfusion MRI}
This imaging modality is instrumental in assessing blood flow to the heart muscle, crucial for identifying regions with insufficient blood supply. It involves injecting a contrast agent and capturing images as the contrast first passes through the heart. Myocardial perfusion MRI is essential for diagnosing and managing conditions like coronary artery disease and ischemic heart disease \cite{xue2020automated,alskaf2022deep}.

\subsubsection{Diffusion-Weighted Imaging (DWI)}
DWI provides a unique perspective by evaluating the diffusion of water molecules within tissues, revealing information about tissue microstructure. This technique is particularly useful for assessing areas of acute or chronic myocardial ischemia, as alterations in water diffusion properties can indicate tissue damage \cite{ueda2022deep}.

\subsubsection{4D Flow MRI}
4D Flow MRI is a cutting-edge technique that visualizes the velocity and direction of blood flow in three dimensions over time. This comprehensive approach to blood flow analysis is crucial for assessing abnormalities in blood vessels, such as aortic aneurysms, aortic dissections, and congenital heart defects \cite{berhane2022deep}.

\section{AI in Individual Imaging CVD  Analysis}
\label{heartmethod}

In this section, we focus on AI pipelines for individual CVD analysis across both non-vessel and vessel cardiac structures. 
Our goal is to summarize key cardiac diseases, the imaging modalities used to evaluate them, and the AI methodologies that support their diagnosis and management. 
%For each disease category, we emphasize the clinical motivation, why accurate segmentation, functional quantification, stenosis detection, or perfusion assessment directly matters for patient care. 
%These tasks influence critical decisions such as determining stroke risk in atrial fibrillation, evaluating ventricular dysfunction in heart failure, identifying ischemia and infarction in myocardial disease, and guiding revascularization strategies in coronary artery disease. 
By linking AI methods to concrete clinical needs, this section highlights how advanced imaging analytics can improve diagnostic precision, reduce variability, and ultimately enhance cardiovascular outcomes.
 
% In this section, we focus on AI pipelines for individual CVD analysis, targeting both non-vessel and vessel cardiac structures. Our aim is to outline prevalent cardiac diseases and their analysis using AI technologies in cardiac imaging. 
% We provide an overview of key diseases, the data modalities used for their study, and recent AI methodologies in this area. 

%The application of representative AI models in cardiac imaging is depicted in Figure \ref{fig_aimodel}, offering insights into the integration of AI in cardiac health and disease management.

%-----------------------------------
\begin{figure*}[t] % use t 
\centering
\includegraphics[width=0.98\linewidth]{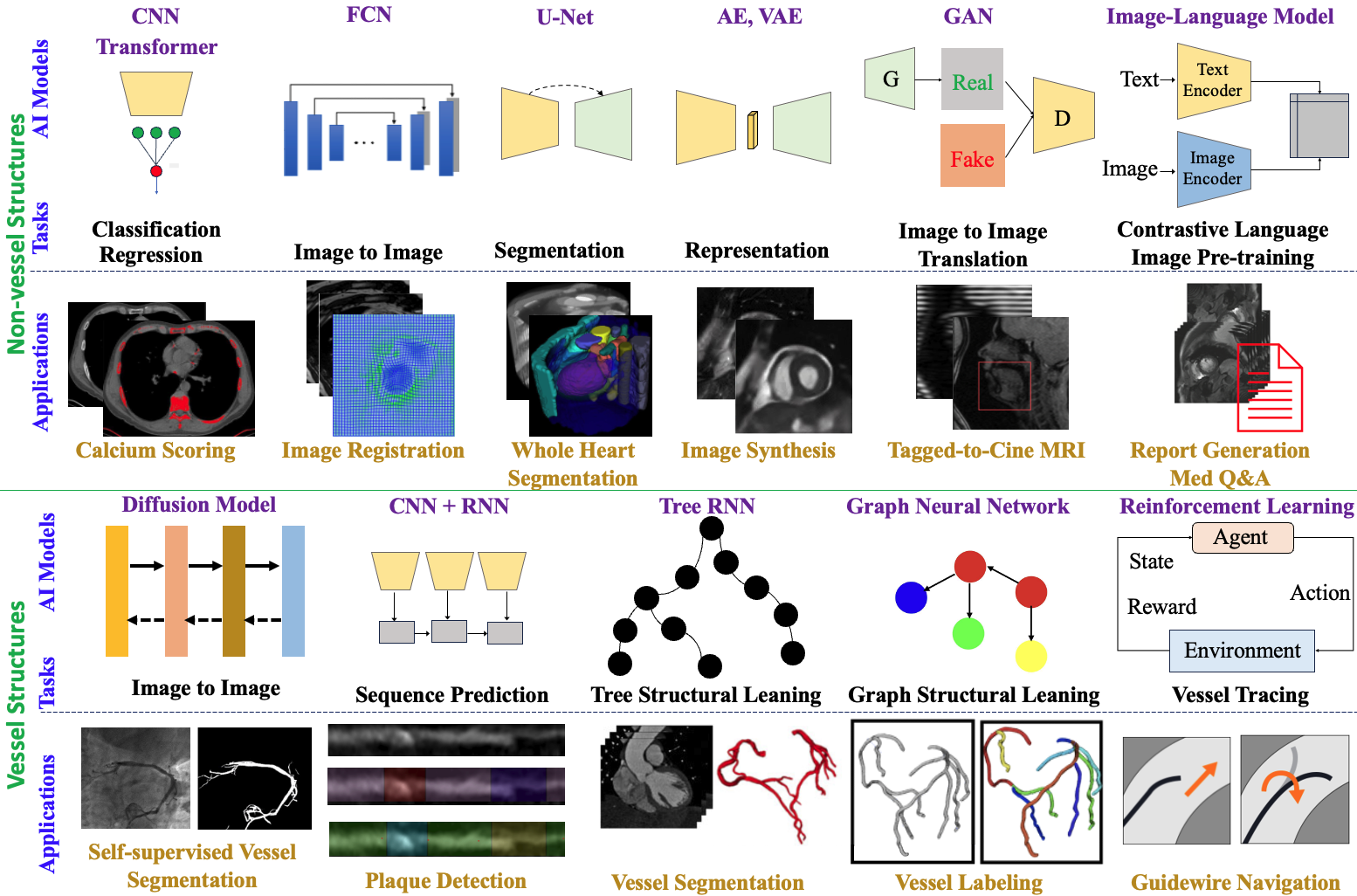}
\caption{\it 
Examples of recent representative AI models for  CVD Analysis.
\textbf{Top (Non-vessel).} \underline{Examples of Networks and Tasks:} Convolutional neural network (CNN) \cite{lecun2015deepronneberger2015u}, Transformer \cite{vaswani2017attention}, Fully convolutional networks (FCN) \cite{long2015fully}, 
U-Net \cite{ronneberger2015u}, Autoencoders (AE) and variational autoencoders (VAE) \cite{ehrhardt2022autoencoders}, Generative adversarial networks (GAN) \cite{goodfellow2020generative}, and Contrastive Language Image Pre-training (CLIP)  \cite{zhao2023clip}.
\underline{Examples of Applications:} Calcium Scoring \cite{isgum2012automatic}, Image Registration \cite{sheikhjafari2022unsupervised}, Whole Heart Segmentation \cite{bui2020simultaneous}, Image Synthesis \cite{jaen2022synthetic}, Tagged MRI to Cine MRI Transform \cite{liu2021dual}, and  Report Generation \cite{litjens2019state}. 
\textbf{Bottom (Vessel).} \underline{Examples of Networks and Tasks:} Diffusion Model \cite{ho2020denoising}, CNN$+$Recurrent Neural Networks (RNN) \cite{donahue2015long}, Tree-structured RNN \cite{lu2023attention},
Graph Neural Network (GNN) \cite{wu2020comprehensive}, and  Reinforcement Learninging (RL) \cite{wang2023deep}.
\underline{Examples of Applications:} Self-supervised Vessel Segmentation \cite{kim2023diffusion}, Vessel Stenosis Detection \cite{zreik2018recurrent},
Vessel Segmentation \cite{lu2023attention}, Vessel Labeling \cite{yang2020cpr} and 
Guidewire Navigation \cite{kweon2021deep}. 
}
\label{fig_aimodel}
\end{figure*}
%----------------------------

%++++++++++++++++++++++++++++++++++++++++++++++++
\subsection{AI Methods}

% CNN Transformer, FCN, UNet, AE, VAE, GAN CLIP, Diffusion Model, RNN, Graph Neural Network, RL 

In this section, we review the most widely used AI models for cardiovascular disease (CVD) analysis. An overview of the models and their applications is shown in Figure \ref{fig_aimodel}.
Convolutional Neural Networks (CNNs) are foundational for image analysis tasks due to their ability to recognize spatial patterns \cite{lecun2015deep}. 
In CVD analysis, CNNs are used for tasks like heart disease classification, plaque detection in blood vessels, and image classification for disease diagnosis. They excel in analyzing echocardiograms, MRI, and CT scans to identify abnormalities and quantify disease severity. 
%Transformer \cite{vaswani2017attention} uses the self-attention mechanism, excelling in handling sequential data and long-range dependencies. 
Transformers can analyze time-series data from ECG signals to detect arrhythmias and other heart rhythm disorders \cite{vaswani2017attention}. 
They are also useful in analyzing 3D imaging data, enabling more accurate analysis of complex cardiac structures by capturing spatial relationships in the heart anatomy.
Fully Convolutional Networks (FCNs): FCNs are designed for pixel-level predictions, making them ideal for tasks requiring fine-grained segmentation \cite{long2015fully}, such as identifying heart chambers, vessel walls, and plaque deposits in imaging. They are widely used in CVD to delineate cardiac structures and assess the impact of cardiovascular diseases on heart anatomy in MRI and CT scans.
% UNet \cite{ronneberger2015u} is a popular architecture for medical image segmentation due to its skip connections that maintain spatial information across network layers. 
UNet is frequently used to segment heart structures \cite{ronneberger2015u}, such as the myocardium and ventricles, and to distinguish tissue types, allowing for more accurate assessment of heart health and disease progression in imaging modalities like MRI. 
% The U-Net architecture, introduced by \cite{ronneberger2015u}, is a prominent model in medical image segmentation.
% Its popularity stems from its flexibility, efficient design, and proven success across various imaging modalities \cite{li2018h}. 
% U-Net comprises two main components: the Encoder and the Decoder. The Encoder path captures semantic and contextual features through downsampling and convolutional blocks, while the Decoder path uses convolutional and upsampling blocks to gradually increase spatial resolution and facilitate pixel-wise classification. 
% U-Net's defining feature is its skip connections, which bridge each level of the Encoder to the corresponding Decoder level, ensuring the transfer of high-resolution contextual details. This feature promotes the combination of low-level details with high-context information for precise localization.  Since its inception, U-Net has established itself as a benchmark in medical image segmentation, inspiring various advanced derivatives that enhance its original design. For a comprehensive review of U-Net variants, see \cite{azad2022medical}. 
Autoencoders (AE) are used for feature extraction, data denoising, and dimensionality reduction \cite{bank2023autoencoders}. 
In CVD analysis, they are helpful for compressing complex imaging data, enabling more efficient storage and analysis. AEs are also used in unsupervised learning to extract critical features from cardiac images or ECG data, supporting downstream tasks like anomaly detection in imaging or rhythm analysis.
Variational Autoencoders (VAEs) extend autoencoders for probabilistic data generation, making them useful for generating synthetic data in CVD research. They can create realistic variations of heart images, which is helpful for augmenting training datasets and improving model robustness. VAEs are particularly useful when working with rare cardiac conditions by generating synthetic examples to balance datasets.
Generative Adversarial Networks (GANs) \cite{goodfellow2020generative} are powerful for creating synthetic but realistic images, supporting CVD analysis by augmenting datasets with realistic heart images. GANs can also be used to translate images between modalities, such as generating MRI-like images from CT scans, enhancing data compatibility and model performance when multi-modality data is limited. GANs also play a significant role in medical image segmentation \cite{creswell2018generative,chen2022generative,xun2022generative,xu2020contrast}. 
Contrastive Language-Image Pretraining (CLIP) aligns images with text descriptions, enabling applications in CVD where imaging findings can be associated with clinical notes \cite{zhao2023clip}. In practice, CLIP allows for cross-modal retrieval, enabling clinicians to search image databases using textual descriptions or retrieve similar cases, streamlining diagnostic workflows and supporting clinical decision-making.
More recently, denoising diffusion models, a notable subset of generative models, have recently attracted significant attention in the field of deep learning \cite{yang2023diffusion}. They have demonstrated remarkable utility across a wide range of applications, especially in the enhancement of medical image segmentation \cite{kazerouni2023diffusion}. 
These models have demonstrated their effectiveness in generating high-quality segmented images, further expanding the possibilities in medical imaging analysis.
In CVD, they can generate diverse and realistic images to simulate disease conditions or augment datasets. 
This is especially valuable for rare diseases or creating data variations that represent disease progression, ultimately improving model training.
Recurrent Neural Networks (RNNs) \cite{shi2015convolutional} are designed for sequential data and are applied in CVD analysis for time-series data \cite{azad2019bi} like ECG signals. 
RNNs are ideal for detecting arrhythmias, analyzing heart rhythm abnormalities, and predicting the progression of heart conditions over time by analyzing patient histories and longitudinal data. 
Graph Neural Networks (GNNs) \cite{zhou2020graph} are suited for data that can be represented as graphs, making them ideal for modeling the connectivity of heart and vascular structures. 
In CVD, GNNs can analyze relationships between cardiac segments or vessel pathways \cite{wu2019automated}, which supports tasks like vessel segmentation \cite{kong2020learning}, detecting aneurysms, and localizing disease within complex vascular networks \cite{gao2020learning}.
Reinforcement Learning (RL) \cite{sutton2018reinforcement} is used in CVD for optimizing sequential decision-making, such as developing personalized treatment plans based on predicted patient responses, guidewire navigation \cite{kweon2021deep}, and image registration \cite{wang2023deep}. 
RL can simulate potential treatment outcomes for different interventions, allowing clinicians to tailor therapy plans and improve patient outcomes in areas like heart failure management or post-surgical recovery planning.

%------------------------------------------------
\begin{figure*}[t] % use t for all the figs and tabs
\centering
\includegraphics[width=1\linewidth]{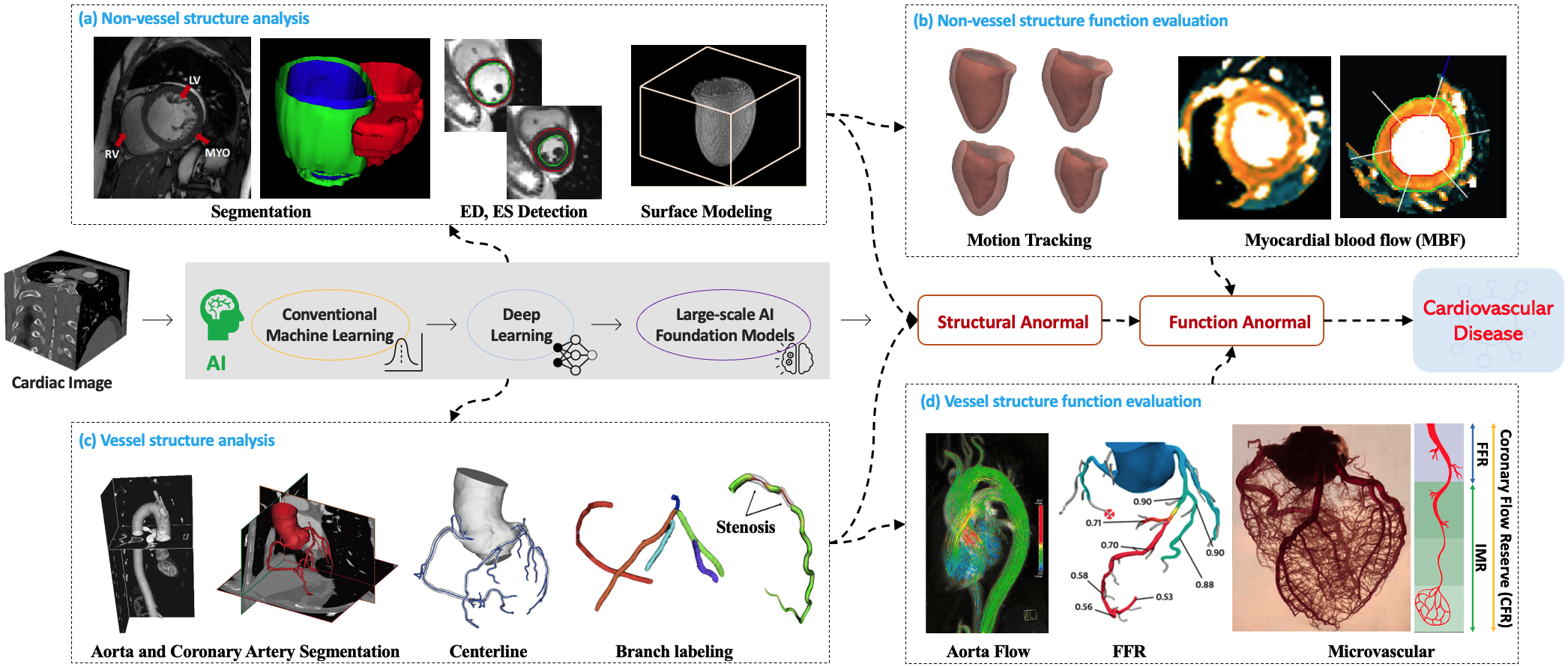}
\caption{
\small
\it  Overview of the cardiac image analysis pipeline and examples of structural and functional imaging. \textbf{Top: Non-vessel.} \underline{(a):} 
%The original image highlighting the Heart Region of Interest (ROI) \cite{zhang2019deep}, including the identification of end-diastole (ED) and end-systole (ES) phases and the segmentation of the Right Ventricle (RV), Left Ventricle (LV), and myocardium (Myo).  Quantification of 6-Segment Division of the Myocardial Mask at ED \cite{zheng2019explainable} and Surface Modeling \cite{bello2019deep}.
Cardiac segmentation \cite{vesal2020fully}, End-diastole (ED) and End-systole (ES) phases detection and Surface Model \cite{meng2022mulvimotion}.
\underline{(b):}  Cardiac Motion Tracking \cite{meng2022mulvimotion}. 
Myocardial blood flow (MBF) maps \cite{jacobs2021automated}. %along with a surface-shaded heart model at end-systole with velocity \cite{bello2019deep}.   
\textbf{Middle:}   Medical images serve as inputs for AI models in the CVD diagnosis pipeline.
\textbf{Bottom: Vessel.} 
\underline{(c):} Segmentation and quantification of vessel structures, with a focus on the Aorta \cite{trentin2015automatic}. Coronary Artery Segmentation \cite{yan2022impact}, centerline extraction using segmentation masks \cite{yan2022impact}, Coronary Artery Branch Labeling based on centerline structures \cite{yang2020cpr}, and detection of Coronary Artery Stenosis \cite{kiricsli2013standardized}. 
\underline{(d):} Evaluation of vessel function. 
Arota flow speed with 3D streamlined visualization \cite{righini2023four}.  
Interpretation of CT-FFR results showing significant hemodynamic impact \cite{baessato2020incremental}, and Coronary Microvascular \cite{clarke2020invasive}. 
Physiological measurements: CFR = coronary flow reserve; FFR = fractional flow reserve; IMR = index of microvascular resistance.
}
\label{fig_ovpp}
\end{figure*}
%------------------------------------------------

%++++++++++++++++++++++++++++++++++++++++++++++++
\subsection{AI Pipelines for Cardiac Image Analysis} 

%^We review  the cardiac image analysis pipeline for both non-vessel and vessel structures. These structures share similarities in their analysis techniques, yet are often studied separately. Existing literature typically focuses on one or the other, neglecting the interplay between them.
%Our survey aims to   emphasize the interconnectedness of non-vessel and vessel structures in the context of  CVD  analysis. We will explore and discuss how these relationships are integral to a comprehensive understanding of CVD. 

Our review delves into the cardiac image analysis pipeline, encompassing both non-vessel and vessel structures, as depicted in Figure \ref{fig_ovpp}. This figure provides a comprehensive overview of the pipeline stages, including segmentation, feature extraction, and registration, alongside specific examples. 
Although these structures share analytical techniques, they are often examined separately in existing literature, leading to an oversight of their interconnected dynamics. 
This separation creates a gap in understanding, as the focus typically remains on one type of structure without acknowledging their interrelation.
In our survey, we emphasize the intertwined nature of non-vessel and vessel structures in CVD analysis. We explore the significance of their interaction in achieving a complete understanding of CVD. This approach offers valuable insights into the intricacies of cardiac health and disease, highlighting the importance of considering both structures in tandem for comprehensive cardiac analysis. 
%By detailing these tasks and their clinical relevance, the section shows how AI can reduce manual workload, improve diagnostic consistency, and help clinicians detect disease earlier and with greater precision.

%\subsubsection{AI-based Segmentation}
% , following the pipeline outlined in 

%As shown in  Figure \ref{fig_ovpp}, cardiac image segmentation is vital in a myriad of medical applications, serving as the foundational step for identifying and analyzing crucial anatomical structures \cite{peng2016review}. 
%It segments the image into regions with distinct semantic meaning, enabling the extraction of various quantitative measures. 
%For non-vessel structures, this includes metrics like the volumes and ejection fractions of the left and right ventricles (LV and RV), the size of atria, as well as myocardial mass and wall thickness \cite{chen2020deep}. 
%For vessel structures, it facilitates the analysis of coronary arteries, including the centerline of coronary arteries, plaques, and more. This segmentation is instrumental in providing comprehensive insights for both diagnosis and treatment. 

As illustrated in the left panel of Figure \ref{fig_ovpp}, the pipelines start with cardiac image segmentation, which plays a crucial role in numerous clinical applications \cite{zhuang2019evaluation,cetin2023attri}. 
This process is key to identifying and analyzing essential anatomical structures \cite{peng2016review}, as it divides the image into regions with distinct semantic meanings and facilitates the extraction of various quantitative metrics. 
In the context of non-vessel structures, segmentation aids in calculating important metrics such as the volumes and ejection fractions of the left and right ventricles (LV and RV), atrial sizes, myocardial mass, and wall thickness \cite{chen2020deep}. 
For vessel structures, segmentation is pivotal for analyzing aspects of coronary arteries \cite{tsai2024uu}, including centerline extraction \cite{fu2023prior}, plaque identification, and other vascular features. Such detailed segmentation provides invaluable insights for both the diagnosis and treatment planning processes.

As illustrated in the right panel of Figure \ref{fig_ovpp}, following segmentation, the AI system engages in feature extraction for precise quantification. This stage involves measuring physical heart parameters like chamber volumes, wall thickness, and myocardial mass \cite{peng2016review}.
Utilizing the segmented structures and the features derived, AI models then conduct functional analysis. This encompasses the calculation of ejection fraction, wall motion, and blood flow dynamics, crucial indicators of the heart's efficiency and overall functioning \cite{peng2016review}. By integrating segmented images, extracted features, and quantification metrics, the AI system forms a comprehensive assessment of cardiac structure and function, invaluable for diagnosing cardiac conditions and formulating treatment plans. 
The details of methods in Figure \ref{fig_ovpp} are introduced in Section \ref{nonvessel} for non-vessel structures and Section \ref{vessel} vessel structures.
For clinical utility, the AI pipeline presents results in an easily interpretable format, often through 3D models or quantitative maps, compiled into reports for cardiologists' evaluation. Additionally, it incorporates clinician feedback to refine models, enhancing accuracy and adaptability to new data and insights. 
This AI-driven approach in cardiology transforms complex imaging data into practical medical information, potentially revolutionizing the diagnosis and treatment of heart diseases.

In the pipeline, the image registration also plays a critical role in analyzing heart structures and diseases, especially when merging data from different imaging modalities or tracking disease progression over time \cite{khalil2018overview,zakeri2022probabilistic,zakeri2023dragnet}. 
Accurate registration is essential for mapping structural changes precisely and is indispensable in longitudinal studies, treatment planning, and assessing therapeutic efficacy \cite{makela2002review}. 
The task of cardiac image registration is particularly challenging due to the significant variability in cardiac shape and motion among individuals \cite{toth20183d}. It often involves aligning heart images captured at different times or under varying conditions, such as different phases of the cardiac cycle or using diverse imaging techniques like CT, MRI, or ultrasound \cite{chen2025survey}. 
Given the dynamic nature of the heart and the influence of respiratory movements, non-rigid registration is crucial for cardiac images \cite{sheikhjafari2022unsupervised}. This approach is key to understanding disease progression or treatment effects by comparing temporal image sequences. 
Advancements in AI models are vital for enhancing registration accuracy and efficiency. The development of real-time registration systems is particularly significant for their potential application in surgical or interventional settings \cite{wang2023deep}, promising to significantly improve patient outcomes and procedural precision in cardiac care \cite{lara2022deep}.

\subsection{Non-vessel Structures}
\label{nonvessel}

In this section, we provide a summary of CVD analysis focusing on the Non-vessel structures.

\subsubsection{Atria and Atrial Diseases}

In this section, we introduce the main structure and function of the atria, as well as AI technologies used for analyzing related diseases.

\paragraph{Atria}
The atria, consisting of the left and right atrium, are the upper chambers of the heart responsible for receiving blood, as depicted in Figure \ref{overview} (a). The right atrium collects blood returning from the body, and the left atrium receives oxygenated blood from the lungs. Both chambers play a crucial role in channeling blood into the ventricles. This section focuses on common diseases that impact the atria and examines the application of AI technologies in their analysis and diagnosis.

\paragraph{Atrial Diseases}
%\noindent\textbf{Atrial Diseases.}  
We examine several common diseases related to the atria \cite{wang2022deep,ter2023juvenile}.  
A key condition, {Atrial fibrillation (AF)}, is a prevalent cardiac arrhythmia marked by irregular and rapid atrial electrical activity \cite{raicea2021giant,wang2022current}. 
Instead of regular contractions, the atria fibrillate or quiver, resulting in an irregular heartbeat, leading to symptoms like palpitations, shortness of breath, fatigue, and dizziness \cite{makynen2022wearable}. Crucially, AF heightens the risk of stroke due to potential blood clot formation in the atria \cite{raghunath2021deep}.
For detecting AF-related abnormalities, primary imaging techniques include ECG and cardiac MRI. ECG is preferred for its real-time imaging, while cardiac MRI, especially Late Gadolinium Enhancement (LGE) MRI, offers high-resolution images vital for assessing structural changes and fibrosis. 
Deep learning algorithms trained on extensive ECG datasets can effectively identify AF patterns \cite{raghunath2021deep}, achieving high accuracy that often matches or exceeds human experts \cite{murat2021review}. As shown in Figure \ref{overview} (a), the AI analysis pipeline for AF ablation involves tasks like LA cavity and wall segmentation, scar segmentation, quantification, and applications such as locating ablation gaps from LGE MRI \cite{li2020atrial}.
{Atrial Enlargement (AE)}, another prevalent atrial condition, involves the enlargement of one or both atria, often resulting from chronic issues like hypertension and heart valve diseases \cite{hsu2022machine}. 
AE is linked to an increased risk of arrhythmias and heart failure \cite{jiang2020detection}. 
Various imaging modalities, including ECG, echocardiography, cardiac MRI, and CT scans, are employed to detect and assess AE. 
ECG stands out for its real-time imaging and widespread availability, effectively identifying atrial size and function. Doppler echocardiography complements ECG by providing detailed hemodynamic data. 
Cardiac MRI and CT scans are valued for their high spatial resolution, enabling precise quantification of atrial volume with heightened sensitivity and specificity. 
Deep learning technologies, particularly Convolutional Neural Networks (CNNs), have been trained on extensive datasets to recognize ECG patterns indicative of AE \cite{jiang2020detection}. 
These models offer clinicians a rapid and accurate preliminary diagnosis based on ECG data. 
%The quantification of AE involves measuring atrial dimensions, volume, and area, as well as assessing shape descriptors and functional parameters like atrial emptying fractions.
While MRI and CT provide high specificity and sensitivity for these measurements, echocardiography may encounter challenges related to image quality and variability among operators \cite{kuchynka2015role}. The integration of these imaging techniques with AI advancements offers a powerful toolkit for the early detection and quantification of atrial enlargement.

\subsubsection{Ventricle and Ventricular Diseases}

%The heart is a vital organ responsible for pumping blood throughout the body. It consists of four chambers: two atria and two ventricles. 

In this section, we introduce the main structure and function of the ventricle, as well as AI technologies used for analyzing related diseases.

\paragraph{Ventricle} 
The heart ventricles, comprising the lower two chambers, are vital components of the heart's anatomy, as illustrated in Figure \ref{overview} (b). The left ventricle, recognized as the heart's largest and strongest chamber, plays a critical role in receiving oxygen-rich blood from the left atrium and pumping it into the systemic circulation. This function is essential for supplying oxygen and nutrients to the body's organs and tissues.
On the other hand, the right ventricle is responsible for receiving oxygen-depleted blood from the right atrium and directing it into the pulmonary circulation \cite{pantelidis2023deep,zhao2022deep}. 
%In this section, we explore various diseases related to the ventricles and delve into how AI technologies are being utilized to analyze these conditions \cite{shoaib2023overview}. This discussion will highlight the intersection of advanced technological applications and cardiology, demonstrating the impact of AI in the understanding and treatment of ventricular diseases. 

\paragraph{Ventricular Diseases} 
%\noindent\textbf{Ventricular Diseases.}
We delve into various anomalies related to the structure and function of the ventricles, each presenting unique characteristics and clinical implications \cite{olaisen2023automatic}. For instance, abnormal heart rhythms can originate from the ventricles, ranging from benign conditions like premature ventricular contractions (PVCs) to more serious disorders such as ventricular tachycardia or fibrillation \cite{madan2022hybrid}. These arrhythmias can compromise the heart's normal pumping function, potentially leading to life-threatening situations \cite{kolk2023machine}. 
In contrast, Ventricular Septal Defect (VSD) is a congenital heart defect characterized by an abnormal opening in the septum dividing the ventricles. This defect allows the mixing of oxygen-rich and oxygen-poor blood, impacting the heart's efficiency. The size and severity of VSDs can vary \cite{yang2023application}.
As indicated in Figure \ref{overview} (b), the Ejection Fraction (EF) is a key measure of the heart's pumping efficiency, particularly of the left ventricle. It is critical for identifying patients at risk of heart dysfunctions such as heart failure \cite{kusunose2021standardize}. EF represents the proportion of blood ejected from the left ventricle with each heartbeat \cite{liu2021deep}. Normally, the left ventricle effectively pumps out over half of its blood content with each beat. A reduced EF signifies impaired heart pumping function \cite{akerman2023automated}. 
A notable recent study introduced a model using Graph Neural Networks (GNNs) to estimate EF from echocardiography videos. This model, EchoGNN, demonstrates EF prediction accuracy comparable to current leading methods and offers crucial explainability, addressing the high variability often seen in observer interpretations \cite{mokhtari2022echognn}.

% For example, {Ventricular Hypertrophy} refers to the thickening of the ventricular walls, primarily caused by the increased workload on the heart \cite{pantelidis2023deep}. 
% It can occur in either the left or right ventricle and is often a result of conditions such as high blood pressure, heart valve disease, or heart failure \cite{zhao2022deep}.
% 室间隔缺损
% 室性心律失常
%\noindent\textbf{Ventricular Arrhythmias.}

%4. Ventricular Septal Rupture: This is a rare but serious complication that can occur after a heart attack. 
%It involves the tearing or rupturing of the ventricular septum, which can cause blood to flow between the ventricles. Ventricular septal rupture requires immediate medical attention and often requires surgical intervention.
%\noindent\textbf{Ventricular Functional Assessment.}

% The objective of ventricle segmentation is to delineate the endocardium and epicardium of the LV and/or RV. These segmentation maps are important for deriving clinical indices, such as left ventricular end-diastolic volume (LVEDV), left ventricular end-systolic volume (LVESV), right ventricular end- diastolic volume (RVEDV), right ventricular end-systolic volume (RVESV), and EF.

\subsubsection{Myocardium and Myocardial Diseases}

In this section, we introduce the main structure and function of the myocardium, as well as AI technologies used for analyzing related diseases.

\paragraph{Myocardium} 
The myocardium, the heart's muscular tissue and its middle layer, is crucial for the organ's function, as illustrated in Figure \ref{overview} (c). Comprising specialized cardiac muscle cells called cardiomyocytes, the myocardium is responsible for the heart's contractions and relaxations, essential for pumping blood throughout the body. Its role in maintaining heart function and ensuring proper blood circulation is fundamental \cite{rinaldi2022invasive}.  
The myocardium receives its supply of oxygenated blood from the coronary arteries, which nourish the myocardial cells with oxygen and essential nutrients. With each heartbeat, the myocardium contracts, effectively propelling blood from the heart chambers into the circulatory system. 
Various diseases can impair the myocardium, affecting its ability to function effectively. 
In the following discussion, we explore several common myocardial conditions and their implications. These diseases illustrate the myocardium's vulnerability and the importance of maintaining its health for overall cardiovascular well-being.

%\noindent\textbf{Myocardial Diseases.} 
\paragraph{Myocardial Diseases} 
We examine several common diseases related to the  Myocardium.
{Myocardial Infarction (Heart Attack)} is a critical myocardial function-related disease, arising when blood flow to a part of the myocardium is obstructed, typically by a clot in the coronary arteries \cite{zheng2019explainable,ibanez20182017}. This blockage can cause damage or death to the affected myocardial area, leading to chest pain, shortness of breath, and potentially severe complications \cite{cho2020artificial}. 
Innovations in deep learning have enabled the use of non-enhanced cardiac MRI to detect and quantify chronic myocardial infarction, potentially reducing reliance on gadolinium contrast injections \cite{zhang2019deep}. 
Additionally, \cite{doudesis2023machine} developed machine learning models that combine cardiac troponin levels with clinical features to evaluate an individual’s myocardial infarction risk. 
Myocardial perfusion MRI, a noninvasive imaging technique, plays a vital role in detecting ischemic heart disease with high accuracy. 
Figure \ref{overview} (c) demonstrates how AI models provide detailed heart blood flow images, crucial for diagnosing conditions associated with impaired blood supply to the heart muscle, thereby aiding in effective cardiac health management.
{Cardiomyopathy}, another prevalent structural myocardial anomaly, encompasses diseases that affect the heart muscle, potentially impacting ejection fraction \cite{vukadinovic2023deep}. 
A notable subtype, Hypertrophic Cardiomyopathy, is a genetic disorder marked by abnormal myocardial thickening, especially in the left ventricle \cite{guo2022artificial}. This thickening can hinder effective blood pumping, manifesting as chest pain, shortness of breath, fainting, and arrhythmias \cite{wang2023deephcd}. Hypertrophic cardiomyopathy is especially significant as a leading cause of sudden cardiac arrest in young people \cite{doi:10.1016/j.jacep.2018.11.004}.

\subsection{Vessel Structures}
\label{vessel}

In this section, we provide a summary of CVD analysis focusing on the vessel structures. 
For each part, we summarize the related diseases and the AI technologies for Coronary Artery Disease (CAD) analysis, including coronary artery segmentation, stenosis detection, and functional simulation.

% atrial fibrillation, acute myocardial infarction, congestive heart failure, and ischemic heart disease

\subsubsection{Aorta and Aortic Diseases}

In this section, we introduce the main structure and function of the aorta, as well as AI technologies used for analyzing related diseases.

\paragraph{Aorta} 
The aorta, the human body's largest artery, is a crucial component of the circulatory system, as depicted in Figure \ref{overview} (d). Arising from the heart's left ventricle, it distributes oxygenated blood throughout the body, providing essential nutrients and oxygen to various organs and tissues \cite{chandrashekar2021deep}. Structurally, the aorta is segmented into the ascending aorta, the aortic arch, and the descending aorta, each playing a distinct role in blood circulation \cite{stonko2023review}. 
This section offers an overview of some common diseases associated with the aorta. Understanding these conditions is essential for comprehending the aorta's critical role in maintaining overall cardiovascular health and identifying potential risks and complications that may arise from aortic disorders.

% 主动脉瘤

% \noindent\underline{Aortic Aneurysm}
% An aortic aneurysm is a localized enlargement or bulging of the aorta's wall. 
% It can occur in different segments of the aorta, including the ascending aorta, aortic arch, or descending aorta. 
% Aneurysms may develop due to factors like high blood pressure, atherosclerosis, connective tissue disorders (such as Marfan syndrome), or trauma. 
% Aortic aneurysms can be life-threatening if they rupture, causing severe internal bleeding \cite{chandrashekar2021deep}.
% 主动脉缩窄
% \noindent\underline{Coarctation of the Aorta} Coarctation of the aorta is a congenital heart defect characterized by a narrowing of the aorta, usually near the site where the ductus arteriosus (a fetal blood vessel) closes after birth. This narrowing restricts blood flow to the lower body and can lead to high blood pressure in the arms and upper body while causing lower blood pressure and diminished pulses in the legs.
% 主动脉炎
% \noindent\underline{Aortitis} Aortitis refers to inflammation of the aortic wall, which can be caused by various factors, including infection (such as syphilis or tuberculosis), autoimmune disorders (like giant cell arteritis or Takayasu's arteritis), or other inflammatory conditions. Aortitis can lead to aortic dilation, aneurysm formation, and an increased risk of dissection.
%\noindent\textbf{Aortic Structural Anomalies.}

%\noindent\textbf{Aortic Diseases.} 
\paragraph{Aortic Diseases} 
We examine several common diseases related to the Aorta.
{Aortic Stenosis}, as outlined in \cite{holste2023severe}, is a condition marked by the narrowing of the aortic valve, impeding blood flow from the left ventricle to the aorta. 
Common in adults due to valve calcification or age-related degeneration, it can cause chest pain, shortness of breath, fatigue, and fainting \cite{elvas2023ai}. 
The evaluation of heart function is critically enhanced through the analysis of hemodynamic flow parameters \cite{ramaekers2023clinician, zhuang2021role}. 
Four-dimensional (4D) flow MRI, a significant advancement in cardiovascular imaging, is extensively discussed in \cite{bissell20234d}. 
This technology enables the detailed study of flow dynamics in vivo, essential for understanding and quantifying the parameters relevant to aortic disease \cite{juffermans20224d}. 
The importance of discerning these flow dynamics lies in their potential to contribute to or worsen vascular diseases, as detailed in \cite{catapano20204d}. Notably, the innovative 4DFlowNet, introduced in \cite{ferdian20204dflownet}, demonstrates the capability to enhance the spatial resolution of 4D flow MRI, showing its effectiveness in actual patient data. 

\subsubsection{Coronary Arteries and Coronary Artery Diseases}

In this section, we introduce the main structure and function of the coronary arteries, as well as AI technologies used for analyzing related diseases.

\paragraph{Coronary Arteries}
Coronary arteries, as illustrated in Figure  \ref{overview} (e),    play a crucial role in the human circulatory system by delivering oxygen-rich blood to the heart muscle \cite{awad2014public,li2001magnetic,pennell1993magnetic,de2023framing,iannattone2020artificial}. 
These arteries branch out from the aorta and envelop the heart, forming an extensive network of vessels essential for cardiac function. 
Diseases affecting the coronary arteries can lead to myocardial ischemia, a condition characterized by reduced blood flow to the heart muscle \cite{smirnova2023spontaneous,krittanawong2021machine,lionakis2022spontaneous}. Chronic ischemia or its associated complications can progressively impair the heart's pumping efficiency, posing significant health risks \cite{d2023deep}. Early detection and timely medical intervention are crucial for conditions related to coronary arteries, as they can be life-threatening if left unaddressed \cite{lin2023coronary}.
In the subsequent discussion, we delve into the realm of AI-driven analysis of diseases associated with coronary arteries. This exploration includes a comprehensive overview of the latest AI methodologies applied in the detection, diagnosis, and management of coronary artery diseases.

\paragraph{Coronary Artery Diseases (CAD)}
% 冠状动脉瘤
% \noindent\textbf{Coronary Artery Aneurysm} is a bulge in a section of a coronary artery. This bulging can impede blood flow and increase the risk of clots or rupture.
%Symptoms of a heart attack include severe chest pain, shortness of breath, sweating, nausea, and pain radiating to the left arm or jaw. 
CAD, the most prevalent form of heart disease, arises when coronary arteries are narrowed or blocked due to fatty plaque buildup \cite{wahab2023developing}. This condition can significantly reduce or completely obstruct blood flow to the heart, leading to angina, manifested as chest pain or discomfort. 
CAD may also precipitate a heart attack, or myocardial infarction, where a part of the heart muscle sustains damage or death due to insufficient blood supply \cite{zhang2022artificial}.
Advanced stages of CAD often necessitate invasive treatments like Percutaneous Coronary Intervention (PCI) or coronary angioplasty, which involve widening narrowed arteries using a balloon device and potentially placing a stent to keep the artery open. In more severe cases with multiple blockages, Coronary Artery Bypass Grafting (CABG) surgery might be performed, creating alternative pathways for blood flow using vessels from other parts of the body \cite{gautam2022current}. Timely and effective management of CAD is crucial. Below, we discuss significant research challenges associated with CAD.
Over the years, AI techniques have become extensively utilized in the field of Coronary Artery Disease detection \cite{alizadehsani2021coronary}. 
For example, {Coronary Artery Stenosis} is 
a predominant structural concern in coronary arteries. The detection, characterization, and monitoring of stenosis severity and plaque development are vital, as these conditions can lead to serious events like heart attacks \cite{zreik2018recurrent}. Medical imaging is central to these tasks \cite{mezquita2023clinical,li2001magnetic,van1995magnetic,pennell1996assessment,pennell1993magnetic,dhawan2004role,botnar2000noninvasive,van1999magnetic}. Notably, interpretations of coronary computed tomography angiography (CTA) by less experienced clinicians often overestimate the severity of stenosis compared to expert analyses \cite{lin2022deep}. AI solutions present a promising alternative. 

Coronary Flow Reserve (CFR) is an essential measure in cardiology, gauging the coronary arteries' capacity to augment blood flow in response to increased cardiac demand, as depicted in Figure \ref{fig_ovpp} (d). 
This metric provides an overarching view of the entire coronary circulatory system's flow capacity, encompassing both large and small vessels. 
It is particularly useful in evaluating patients who exhibit ischemic symptoms without significant stenosis, helping to guide decisions on revascularization or medication management. 
CFR's estimation can be achieved non-invasively through MRI and PET, etc, offering a safer alternative for certain patient groups \cite{fawaz2023invasive}. 
Typically, CFR is assessed through invasive coronary catheterization, where blood flow velocity is measured at rest and during induced hyperemia using specialized Doppler guidewires or pressure sensors.
% It's essential for assessing the health of both the macrovascular and microvascular components of the coronary circulatory system, measuring how much blood flow can increase from its resting state to meet the heart's increased oxygen needs.  
% A normal CFR value is above 2.0, indicating adequate coronary dilation for increased blood flow. 
% Reduced CFR suggests impaired flow capacity, often due to coronary artery disease (CAD), microvascular dysfunction, or other abnormalities. 

Fractional Flow Reserve (FFR) is an index for quantifying blockage severity in coronary arteries, typically determined through invasive catheter-based methods, as illustrated in Figure \ref{overview} (e). 
However, modern approaches are exploring non-invasive FFR estimation, using anatomical data from CT scans of the heart and coronary arteries. 
Despite their promise, these methods, mainly physics-based models, face challenges due to high computational demands, limiting their routine clinical application.
With the integration of computational AI, there's a growing interest in developing non-invasive FFR estimation techniques from medical imaging modalities like coronary computed tomography angiography (CCTA). These AI-driven methodologies are aimed at circumventing the need for invasive procedures, making CAD assessment more patient-friendly and efficient. As seen in Figure \ref{fig_ovpp} (c, d), a typical AI-powered FFR analysis workflow includes segmentation of coronary arteries within CTA images to delineate their anatomical structure \cite{zheng2011machine,kong2020learning,zeng2022imagecas}. This process is often supplemented with automated branch labeling systems \cite{li2023automated}, enhancing lesion localization accuracy in coronary arteries \cite{wu2019automated, yang2020cpr, zhao2023agmn,li2020segmentation}. 
The culminating step involves leveraging deep learning models, trained on the segmented and labeled structural and functional features, to predict FFR. These AI models present a compelling alternative to traditional physics-based methods, offering potentially quicker and equally accurate FFR estimations \cite{itu2016machine,gao2020learning}.

%\textbf{Coronary Arteries Segmentation}
%This method integrates cutting-edge deep learning algorithms with medical imaging data to estimate FFR values. 

%Coronary artery segmentation plays a pivotal role in the diagnosis and management of coronary artery diseases. 
%With advancements in AI techniques, deep learning-based automated segmentation has shown promising results, reducing the time and effort required while increasing accuracy and reproducibility \cite{zeng2022imagecas}. 

\paragraph{Coronary Microvascular Disease (CMVD)} 
CMVD disease within the coronary circulatory system, particularly affecting the microvasculature, can lead to clinical events in patients without obvious epicardial coronary stenosis \cite{ayub2020coronary}. CMVD specifically targets the heart's small blood vessels or microcirculation, often resulting in symptoms like chest pain and shortness of breath, though it can also be asymptomatic. Unlike obstructive coronary artery disease, which impacts larger vessels, CMVD involves reduced blood flow through these microvessels (See Figure \ref{fig_ovpp} (d)). 
Traditional angiography typically fails to detect CMVD, as it mainly visualizes larger coronary arteries, necessitating alternative diagnostic approaches.
% in Microvascular Assessment
The Index of Microcirculatory Resistance (IMR) serves as a key invasive metric for directly evaluating coronary microvascular function. 
This index is obtained using specialized pressure and temperature-sensitive wires, similar to those used in FFR measurements. 
It's crucial to recognize that high FFR values, typically above 0.80, do not entirely rule out the risk of future clinical events. Patients with high FFR readings may still experience clinical issues, emphasizing the need for comprehensive assessments beyond FFR \cite{berry2014fractional}.
IMR is particularly insightful for diagnosing microvascular dysfunction, proving valuable in both stable patients and those with acute or recent MI \cite{zhao2023early}. 
%Despite its utility, the medical field still seeks reliable and non-invasive quantitative techniques for diagnosing CMVD. 
Developing such methods is essential for selecting appropriate treatments for coronary artery disease and improving coronary microcirculation outcomes \cite{gao2022clinical}.

\subsection{Connections of Non-vessel and Vessel Structures}
%Connections Non-vessel and Vessel Structures for the cardiac function

%\xin{ LV seg and vessels}
%\xin{others?}
%\xin{ read the CAD feature paper}

The heart's functionality is intricately dependent on both its non-vessel and vessel structures.   Understanding the interplay between these two types of structures is pivotal for a holistic comprehension of cardiac function, specifically how structural integrity and blood supply interact.
%As illustrated in Figure \ref{fig_relation} (Top), 
The myocardium's dependency on a continuous supply of oxygen and nutrients, primarily delivered by the coronary arteries, is a key aspect of cardiac function. 
Disruptions in this supply, such as from atherosclerosis or thrombosis in the coronary arteries, can result in myocardial ischemia or infarction, impairing the heart's pumping efficiency \cite{li2023multi}. 
%
%Conversely, changes in non-vessel structures, like ventricular hypertrophy or valvular diseases, can influence the heart's blood demand and workload, subsequently affecting coronary circulation. 
%For instance, ventricular hypertrophy heightens the myocardium's oxygen needs, which can lead to an imbalance if the coronary arteries are unable to meet this increased demand. 
There exists a dynamic feedback loop between myocardial performance and coronary vessel health. 
For example, myocardial damage from infarction can alter blood flow dynamics, impacting cardiac output and the stress on coronary vessels \cite{fukushima2020patient}. 
%The cardiac cycle involves myocardial contraction and relaxation, as well as valve operation, all of which are vital for effective blood propulsion through the coronary vessels and into systemic circulation.
Structural or functional abnormalities in these non-vessel components can hamper blood circulation efficiency, thereby affecting overall cardiac function \cite{slomka2021quantitative}.
%As shown in Figure \ref{fig_relation} (Bottom), 
% The myocardial computed tomographic perfusion (CTP) derived fractional flow reserve (CTP-FFR) models have been initially established \cite{pan2016clinical,gao2023novel}. 
% The models demonstrate superior diagnostic capabilities for coronary artery disease (CAD), particularly in cases of mild to moderate stenosis.  
A recent work developed a multiscale, patient-specific model to enable blood flow simulation from major coronary arteries down to myocardial tissue \cite{papamanolis2021myocardial}. 
Specifically, a stand-alone coronary model and an integrated coronary-myocardium coupled model were developed and examined with the objective of simulating myocardial perfusion under both healthy and pathological conditions.
%The patient's vasculature is initially segmented from coronary computed tomography angiography data and then extrapolated to the arteriole level using a network of synthetic trees. 
Then, the CT-FFR blood flow is modeled by integrating a coronary artery model with a single-compartment Darcy myocardium model for the joint hyperemic Myocardial Blood Flow (MBF) for the coupled model. 
Finally, the predicted and true perfusion maps are compared. This work represents an instance of a computational model simulating blood flow from epicardial coronary arteries to the left ventricle myocardium, applied and validated using human data.
Currently, the application of AI in comprehensively understanding the connection between non-vessel and vessel structures remains limited. 
However, such understanding is critical for diagnosing and treating cardiac diseases, as it aids in identifying the root causes of cardiac dysfunction, informs surgical intervention planning, and supports long-term cardiac health management strategies. 

\subsection{Section Summary}
Section 3 reviews how modern AI methods are applied across the cardiac imaging pipeline for both non-vessel structures (atria, ventricles, myocardium) and vessel structures (aorta, coronary arteries, microvasculature. 
AI supports key tasks such as segmentation, motion tracking, feature extraction, stenosis detection, and physiologic assessment (e.g., FFR, CFR, IMR). 
%The section underscores that these imaging biomarkers directly impact patient outcomes: EF predicts heart failure risk, atrial fibrosis guides ablation, perfusion maps identify ischemia, and noninvasive FFR informs revascularization decisions. 
It also highlights the need for AI models that jointly consider structural and functional information, reflecting the physiological coupling between myocardium and coronary circulation. Overall, Section 3 shows how AI can enhance diagnostic accuracy and clinical decision-making across the spectrum of cardiovascular disease.

% % 冠状动脉痉挛
% \noindent\textbf{Coronary Artery Spasm (CAS)}
% Also known as Prinzmetal's or variant angina, it's caused by the temporary tightening (spasm) of the muscles in the artery wall.

% \begin{figure}[t] % use t for all the figs and tabs
% \centering
% \includegraphics[width=1\linewidth]{fig/cdfGTFFR.png}
% \caption{ DeepFFR use CFD as GT. \cite{gao2020learning}
% }
% \label{fig_branch}
% \end{figure}

%++++++++++++++++++++++++++++++++++++++++++++++++
%\section{Heart Feature}

%++++++++++++++++++++++++++++++++++++++++++++++++
\section{ Population Image-based CVD Analysis}
\label{PopulationIG}

% public health-related

%\xin{heart image relation more topics, life style, smoking, Genomics age gender}

Findings from population-based imaging studies provide a foundation for individual analysis by identifying broader risk factors and disease progression patterns, which can be refined for personal application. 
%As shown in Figure \ref{fig_gwas} (left), 
The connection between individual and population-based studies is in their shared goal of improving cardiovascular health, but they complement each other by bridging individual and group insights. 
Conversely, individual analysis can inform population studies by revealing which subgroups might benefit most from certain interventions, further refining predictive models. 
Together, these approaches enhance our understanding of CVD and support the transition from broad-spectrum therapies to personalized, data-driven care that considers both individual and population health factors.
Thus, our exploration in this section delves into understanding CVDs through detailed population-based imaging studies, utilizing AI and statistical tools.
We address the diversity of CVDs, like coronary artery disease, heart failure, and arrhythmias, each with its own clinical and pathophysiological traits.   
Acknowledging the significant influence of lifestyle and behavioral factors on these diseases, our focus extends to how these elements, along with genetics and age, shape CVD risks \cite{sing2003genes}. 
Population-based study is important to construct temporal Causal Genetic-Imaging-Clinical (CGIC) pathways to examine the causal relationships between genetic predispositions, cardiac imaging features, and clinical outcomes in CVDs \cite{zhao2023heart}. 
This integrative approach is designed to deepen our understanding of CVD etiology and to enhance strategies for prevention, diagnosis, and treatment.

%\subsection{ Causal Pathways from Cardiac Imaging Features to CVDs}

% a comprehensive biomedical research resource encompassing clinical, imaging, and blood biochemistry data

\subsection{Cardiac Imaging Genetics}

The intricate genetic architecture of CVDs has been elucidated through numerous genome-wide association studies (GWAS) over the past decade and a half, highlighting the interplay of genetic and environmental factors in these diseases. 
Studies cataloged in the GWAS Catalog \cite{buniello2019nhgri} reveals that nearly 1,700 studies on 450 CVD-related phenotypes have been conducted. 
This research has advanced our understanding of cardiovascular biology and pathophysiology, uncovering complex interactions between genetic and non-genetic risk factors. 
Overall, these studies underline the importance of continued exploration into the genetic basis of CVDs to enhance prevention, diagnosis, and treatment strategies.
Heritability estimates for CVDs vary across different conditions. 
For instance, Coronary Heart Disease and some arrhythmias like atrial fibrillation have heritability estimates between 40-60\%. 
A study on twins estimated the heritability of atrial fibrillation to be as high as 62$\%$ \cite{christophersen2009familial}, indicating a strong genetic component. Similarly, hypertension shows a heritability of about 30-50\%. While heart failure and stroke also have significant heritability, the extent varies depending on the type and associated conditions.
Case-control GWAS is the most direct method to discover common genetic risk factors underlying disease.  
GWAS initially focused on common complex diseases like myocardial infarction (MI), atrial fibrillation (AF),  CAD, heart failure, stroke, and hypertension. 
For instance, although the heritability of CAD/MI is estimated to be around 40-50\%, only about 30\% of this heritability is explained by the currently identified Single Nucleotide Polymorphisms (SNPs) \cite{aragam2022discovery,tcheandjieu2022large}. 
This suggests that there are still genetic factors of CAD/MI yet to be discovered.
Thus, utilizing AI models allows for exploring intricate relationships between genetic variations and imaging characteristics, facilitating the identification of new associations and the creation of improved predictive models. 
AI algorithms can effectively amalgamate genetic and imaging data, enabling the delineation of distinct disease subtypes grounded in both genetic profiles and imaging attributes \cite{bueno2023qrs}.

\subsection{Lifestyle and Environmental Factors}

The analysis of population-based data illuminates the complexities and prevalence of cardiovascular conditions, highlighting the multifaceted influences on heart health. This includes genetic predispositions, lifestyle choices such as diet and physical activity, and environmental factors, all contributing to cardiovascular risk profiles \cite{munzel2022environmental,bhatnagar2017environmental,powell2022social}. Modern genomics connects genetic markers to heart health changes, while demographic factors like age and gender add another layer to cardiovascular risks. Integrating cardiovascular imaging with data on lifestyle, genetics, demographics, and environmental factors facilitates a holistic view of heart health, enabling tailored and more effective cardiovascular interventions \cite{barratt2023environmental}. 
Environmental influences, ranging from geographic location changes to lifestyle adaptations, air pollution, and social policies, significantly affect cardiovascular health  \cite{munzel2022environmental,bhatnagar2017environmental,powell2022social}. The interplay between the natural environment, including aspects like sunlight exposure and green spaces, and the social environment, characterized by urbanization, pollution, and social networks, shapes lifestyle choices that impact cardiovascular risks. A comprehensive understanding of these interactions is crucial for devising innovative prevention and treatment strategies, addressing the global challenge of cardiovascular diseases. Advancements in AI and imaging genetics are poised to enhance our understanding of these complex relationships, offering promising avenues for research and clinical practice in cardiovascular health.

%--------------------------
\subsection{Interactions between Heart and other Organs}

The integration of AI in analyzing cardiovascular diseases through imaging techniques has significantly advanced our understanding of the heart's connections with other bodily organs and systems. This comprehensive approach enables the identification of complex interrelations, such as the cardiovascular system's impact on neurological health, and vice versa. For instance, AI algorithms are instrumental in analyzing cerebral blood flow and detecting stroke risks, providing insights into the cardiovascular origins of neurological events.
%
%As shown in Figure \ref{fig_gwas} (right), 
A notable study \cite{zhao2023heart} leveraging multiorgan MRI data from a large cohort has illuminated the phenotypic and genetic connections between cardiovascular health and brain attributes. 
This research has identified genomic loci that influence both heart and brain characteristics, suggesting a shared genetic foundation for cardiovascular and neurological conditions. Such findings underscore the potential of heart conditions to contribute causally to brain disorders, offering new perspectives on human health.
AI's role extends beyond mere data analysis; it facilitates the extraction of critical insights from imaging data, encompassing the heart and its systemic interactions. By employing advanced AI techniques and multimodal data fusion, these models enhance the detection and quantification of abnormalities. This not only aids in accurate diagnosis and risk assessment but also enriches treatment strategies for cardiovascular diseases. 
%The amalgamation of AI with extensive patient data, including imaging and clinical records, presents a holistic view of the cardiovascular system's interplay with other organs, thereby promising to improve patient care and outcomes significantly.

\subsection{Section Summary}
In summary for section \ref{PopulationIG},  we discussed how large-scale imaging data is leveraged to understand CVD trends, risk factors, and progression patterns across diverse populations. 
Population-level imaging studies can identify associations between anatomical variations and CVD prevalence, uncovering insights into how risk factors (such as age, gender, ethnicity, and lifestyle) impact disease development. 
By analyzing imaging data from broad, heterogeneous cohorts, these studies provide normative baselines and detect atypical presentations, which are invaluable for risk stratification and early intervention. 
Additionally, population-based CVD analysis informs public health policies, enabling targeted prevention efforts and resource allocation tailored to the cardiovascular health needs of specific communities or demographics.

\section{Future Directions}
\label{futuredirection}

%While significant progress has been made in leveraging AI for image-based CVD analysis, there are still several avenues for future research and development. 
Exploring the future of cardiovascular health through AI-driven models anchored in medical imaging, alongside auxiliary data such as genetics, clinical history, and lifestyle factors, opens a path to transformative clinical innovation. 
Medical imaging provides a foundational view of the heart’s structure and function, and when combined with auxiliary data, AI technologies can deliver unprecedented insights into cardiovascular disease (CVD) that extend beyond traditional analysis. 
This integrative approach enables personalized, precise diagnosis and treatment strategies tailored to each patient’s unique profile, making precision medicine a reality in cardiovascular care.

\subsection{Digital Twins (DT) of Heart} 
The goal of creating a digital twin of the heart is to facilitate tailored and precise treatments for CVD \cite{corral2020digital}. 
The process involves collecting a wide array of patient data, including advanced imaging, genetic profiles, and physiological metrics \cite{tao2018digital}. 
These data points are synthesized using sophisticated AI techniques to construct a dynamic, virtual model of the patient's heart, reflecting its detailed anatomy and functions. 
The DT model enables simulations to test various therapeutic strategies, offering insights into the most effective treatments. 
%It's continuously updated with new patient data, allowing for real-time health monitoring and enabling informed clinical decisions to adjust treatments as needed. 
Digital twins of the heart offer significant benefits by enabling healthcare professionals to test and refine treatment strategies in a risk-free environment \cite{coorey2022health}. 
This innovative approach can lead to safer patient care, lower healthcare costs, and improved outcomes. Additionally, digital twins have the potential to spur new therapeutic discoveries and enhance our understanding of cardiovascular diseases \cite{ghatti2023digital}. 
However, their widespread implementation faces hurdles such as ensuring data privacy, meeting computational demands, and achieving model accuracy.

\subsection{AI-based Cardiac Image Generation} 
Digitally Reconstructed Radiographs (DRR) serve as a pivotal technique in medical imaging, simulating X-ray images from 3D medical data such as CT or MRI scans, thereby reducing unnecessary radiation exposure for patients \cite{lo2012extraction}. The integration of advanced deep learning models, including 3D convolutional neural networks, GANs \cite{goodfellow2020generative}, variational autoencoders \cite{kingma2019introduction}, and diffusion model \cite{he2023dmcvr,kazerouni2023diffusion} enhances the reconstruction quality but demands considerable computational resources and extensive training data \cite{maken20232d}. 
The advent of Neural Radiance Fields (NeRF) has brought significant advancements in natural image reconstruction, though their application in medical imaging remains challenging due to the complexity of medical data \cite{mildenhall2021nerf}. Innovations like Medical image NeRFs \cite{corona2022mednerf,hu2023umednerf} are making strides by learning to map radiance values to pixels, revealing intricate details of internal anatomy from 2D images, and offering potential for reducing radiation risks and examination costs, especially in orthopedics and surgery.

\subsection{AI Foundation Models for Heart Health} 
AI is increasingly becoming the primary approach for analyzing CVD through imaging \cite{olawade2024advancements}, aiming to reduce costs and the reliance on invasive procedures \cite{gao2020learning,wu2019automated,kong2020learning}. 
AI algorithms have been developed for tasks such as disease classification, risk prediction, treatment planning, and clinical decision-making in cardiovascular imaging \cite{ma2019iterative}. 
However, most current AI approaches in this field rely on task-specific models \cite{lu2023attention}, which may not fully capture the complexity and diversity of medical data due to their smaller size and task-focused design.
The emergence of AI foundation models, particularly prominent in computer vision and natural language processing \cite{zhou2023comprehensive,sun2023survey,zhou2023foundation,li2023multimodal}, represents a significant shift. 
These large-scale models, with millions or billions of parameters, are capable of handling more complex data and identifying a wide range of patterns and relationships, often achieving state-of-the-art results across various tasks due to their scalability and generalization capabilities \cite{radford2018improving,nguyen2023lvm,wang2023medfmc,wang2023foundation,lin2024robust}.
In cardiovascular health, foundation models have the potential to be adapted or fine-tuned for specific CVD analysis tasks, such as Atrial Fibrillation \cite{li2022medical}, Ischemic Heart Disease \cite{xue2020automated}, and Coronary Artery Diseases \cite{gao2020learning}. 
Adapting these models to the diverse imaging modalities used in CVD analysis presents challenges due to the significant differences from natural images, based on various physics-based properties and energy sources \cite{zhang2023challenges}. 
A potential solution involves training foundation models on both medical and natural images through fine-tuning approaches \cite{li2023llava,ma2024segment}, aiming to provide a robust foundational solution for clinical challenges in CVD. 
This advancement could significantly improve the effectiveness and efficiency of diagnosing and treating CVD, marking a substantial step forward in the field \cite{wang2024u}.

%Furthermore, developing more efficient neural network architectures is crucial for advancing foundation models. Notably, Mamba \cite{gu2023mamba}, a State Space Model, has emerged as a significant innovation for its adeptness at efficiently modeling long-range dependencies in sequential data, attributed to its exceptional memory efficiency and computational speed \cite{xing2024segmamba,ma2024u}.

\subsection{Explainable CVDs Diagnosis Models with LLMs}  
Large Language Models (LLMs) hold significant promise in revolutionizing the diagnosis of CVDs, the leading global cause of mortality. 
By leveraging their vast data processing capabilities, LLMs can substantially improve diagnostic precision \cite{liu2025medchat}. 
However, the opaque nature of these models raises concerns about transparency and trust among medical professionals and patients. 
To address this, it's crucial to incorporate explainability into LLMs, ensuring they not only deliver accurate diagnoses but also provide comprehensible rationales for their decisions \cite{teng2022survey}.  
Enhancing the explainability of LLMs \cite{zhao2023explainability} involves creating or improving interpretability methods tailored to medical diagnostic needs. This step is essential for the models' acceptance and practical application in healthcare. 
Moreover, thorough real-world clinical validations are imperative to confirm the model's dependability and effectiveness across various healthcare applications.

% \subsection{Section Summary}
% In summary for Section \ref{futuredirection}, by explicitly linking each direction with a challenge or gap in the current landscape of AI for CVD analysis, this section can offer a well-rounded vision of why these future directions matter and how they could substantially benefit both the research community and clinical practice. 

%+++++++++++++++++++++++++++++++++++++
\section{Conclusion}
\label{conclusion}

%The integration of AI into cardiovascular imaging has demonstrated its ability to improve diagnostic accuracy, assist in personalized treatment planning, and enable early disease detection. 
%In this paper, we provide a comprehensive overview of the current state and new perspectives of AI in image-based cardiovascular disease analysis. 
%We have explored various aspects of image-based cardiovascular disease analysis, including image modalities, image processing, and disease analysis methods. 
%We also discussed the challenges associated with large-scale data collection, annotation, and standardization and the importance of public datasets and code repositories in facilitating research collaborations and algorithm validation. 
%Furthermore, we have highlighted emerging trends, such as imaging genetics, digital twins and AI foundational models, which hold great potential for advancing the field further.
%In the future, the continued advancements in AI algorithms, hardware capabilities, and imaging technologies will pave the way for more accurate and efficient diagnosis, risk stratification, and treatment planning. 
%Integration with electronic health records, genomics, and other clinical data sources into large-scale AI models will enable a more holistic approach to patient care, facilitating personalized cardiovascular care. 

The integration of AI in cardiovascular imaging is significantly enhancing diagnostic accuracy, personalized treatment planning, and early disease detection in the field of cardiology. 
This paper offers an in-depth examination of AI applications in cardiovascular disease analysis through imaging, covering a range of topics from imaging modalities and processing techniques to disease-specific analysis methods.
Challenges such as the collection, annotation, and standardization of large-scale datasets are discussed, emphasizing the critical role of public datasets and code repositories in promoting research collaboration and algorithm validation. 
The paper sheds light on promising future directions, including imaging genetics, digital twins, and the utilization of AI foundational models, which are expected to drive substantial progress in cardiovascular imaging. 
%Looking ahead, the paper anticipates further advancements in AI algorithms, hardware, and imaging technologies, which will lead to more precise and efficient diagnostic and treatment processes. 
Moreover, the integration of AI with electronic health records, genomics, and other clinical data will facilitate a comprehensive and personalized approach to cardiovascular care, aligning with the evolving landscape of patient-centered healthcare.

\section{ACKNOWLEDGMENTS}
Copyrights of the figures belong to their own authors and/or holders (sources are cited in the captions).
%The author sincerely thanks all readers for taking the time to read the manuscript. 
%If any related work has been overlooked, please do not hesitate to contact us. The feedback will be greatly appreciated and help us considerably improve this survey.
%
%Xin Wang is supported by SUNY at Albany Start-up Grant.
Dr. Zhu's work was partially supported by the Gillings Innovation Laboratory on generative AI and by grants from the National Institute on Aging (NIA) of the National Institutes of Health (NIH), including   1R01AG085581, and RF1AG082938, the National Institute of Mental Health (NIMH) grant 1R01MH136055,  and the NIH grants R01AR082684, and 1OT2OD038045-01. The content of this paper is solely the responsibility of the authors and does not necessarily represent the official views of these institutions.

\section{DISCLOSURE STATEMENT}
The authors are not aware of any affiliations, memberships, funding, or financial holdings that might be perceived as affecting the objectivity of this review.

%% chose style
%\bibliographystyle{asa}
%\bibliographystyle{chicago}

% this style used in Statistical Learning Methods for Neuroimaging Data Analysis with Applications
\bibliographystyle{ar-style3}

% chose bib file
\bibliography{main}

@inproceedings{ronneberger2015u,
  title={U-net: Convolutional networks for biomedical image segmentation},
  author={Ronneberger, Olaf and Fischer, Philipp and Brox, Thomas},
  booktitle={International Conference on Medical image computing and computer-assisted intervention},
  pages={234--241},
  year={2015},
  organization={Springer}
}

@article{kingma2019introduction,
  title={An introduction to variational autoencoders},
  author={Kingma, Diederik P and Welling, Max and others},
  journal={Foundations and Trends{\textregistered} in Machine Learning},
  volume={12},
  number={4},
  pages={307--392},
  year={2019},
  publisher={Now Publishers, Inc.}
}

@article{wu2020comprehensive,
  title={A comprehensive survey on graph neural networks},
  author={Wu, Zonghan and Pan, Shirui and Chen, Fengwen and Long, Guodong and Zhang, Chengqi and Philip, S Yu},
  journal={IEEE transactions on neural networks and learning systems},
  volume={32},
  number={1},
  pages={4--24},
  year={2020},
  publisher={IEEE}
}

@article{zhou2020graph,
  title={Graph neural networks: A review of methods and applications},
  author={Zhou, Jie and Cui, Ganqu and Hu, Shengding and Zhang, Zhengyan and Yang, Cheng and Liu, Zhiyuan and Wang, Lifeng and Li, Changcheng and Sun, Maosong},
  journal={AI Open},
  volume={1},
  pages={57--81},
  year={2020},
  publisher={Elsevier}
}

@article{shen2017deep,
  title={Deep learning in medical image analysis},
  author={Shen, Dinggang and Wu, Guorong and Suk, Heung-Il},
  journal={Annual review of biomedical engineering},
  volume={19},
  pages={221--248},
  year={2017},
  publisher={Annual Reviews}
}

@article{zhou2021review,
  title={A review of deep learning in medical imaging: Imaging traits, technology trends, case studies with progress highlights, and future promises},
  author={Zhou, S Kevin and Greenspan, Hayit and Davatzikos, Christos and Duncan, James S and Van Ginneken, Bram and Madabhushi, Anant and Prince, Jerry L and Rueckert, Daniel and Summers, Ronald M},
  journal={Proceedings of the IEEE},
  year={2021},
  publisher={IEEE}
}

@article{wang2024screening,
  title={Screening and diagnosis of cardiovascular disease using artificial intelligence-enabled cardiac magnetic resonance imaging},
  author={Wang, Yan-Ran and Yang, Kai and Wen, Yi and Wang, Pengcheng and Hu, Yuepeng and Lai, Yongfan and Wang, Yufeng and Zhao, Kankan and Tang, Siyi and Zhang, Angela and others},
  journal={Nature Medicine},
  pages={1--10},
  year={2024},
  publisher={Nature Publishing Group US New York}
}

@article{bank2023autoencoders,
  title={Autoencoders},
  author={Bank, Dor and Koenigstein, Noam and Giryes, Raja},
  journal={Machine learning for data science handbook: data mining and knowledge discovery handbook},
  pages={353--374},
  year={2023},
  publisher={Springer}
}

@article{khalil2018overview,
  title={An overview on image registration techniques for cardiac diagnosis and treatment},
  author={Khalil, Azira and Ng, Siew-Cheok and Liew, Yih Miin and Lai, Khin Wee},
  journal={Cardiology research and practice},
  volume={2018},
  year={2018},
  publisher={Hindawi}
}

@article{makela2002review,
  title={A review of cardiac image registration methods},
  author={Makela, Timo and Clarysse, Patrick and Sipila, Outi and Pauna, Nicoleta and Pham, Quoc Cuong and Katila, Toivo and Magnin, Isabelle E},
  journal={IEEE Transactions on medical imaging},
  volume={21},
  number={9},
  pages={1011--1021},
  year={2002},
  publisher={IEEE}
}

@article{chen2025survey,
  title={A survey on deep learning in medical image registration: New technologies, uncertainty, evaluation metrics, and beyond},
  author={Chen, Junyu and Liu, Yihao and Wei, Shuwen and Bian, Zhangxing and Subramanian, Shalini and Carass, Aaron and Prince, Jerry L and Du, Yong},
  journal={Medical Image Analysis},
  volume={100},
  pages={103385},
  year={2025},
  publisher={Elsevier}
}

@article{sheikhjafari2022unsupervised,
  title={Unsupervised diffeomorphic cardiac image registration using parameterization of the deformation field},
  author={Sheikhjafari, Ameneh and Krishnaswamy, Deepa and Noga, Michelle and Ray, Nilanjan and Punithakumar, Kumaradevan},
  journal={arXiv preprint arXiv:2208.13275},
  year={2022}
}

@article{lara2022deep,
  title={Deep learning-based image registration in dynamic myocardial perfusion CT imaging},
  author={Lara-Hernandez, A and Rienm{\"u}ller, T and Ju{\'a}rez, Ivan and P{\'e}rez, Michaelle and Reyna, Favio and Baumgartner, Daniela and Makarenko, Vladimir N and Bockeria, Olga L and Maksudov, Muzaffar and Rienm{\"u}ller, Rainer and others},
  journal={IEEE Transactions on Medical Imaging},
  volume={42},
  number={3},
  pages={684--696},
  year={2022},
  publisher={IEEE}
}

@article{liu2025medchat,
  title={Medchat: A multi-agent framework for multimodal diagnosis with large language models},
  author={Liu, Philip R and Bansal, Sparsh and Dinh, Jimmy and Pawar, Aditya and Satishkumar, Ramani and Desai, Shail and Gupta, Neeraj and Wang, Xin and Hu, Shu},
  journal={IEEE 8th International Conference on Multimedia Information Processing and Retrieval (MIPR)},
  year={2025}
}

@article{xun2022generative,
  title={Generative adversarial networks in medical image segmentation: A review},
  author={Xun, Siyi and Li, Dengwang and Zhu, Hui and Chen, Min and Wang, Jianbo and Li, Jie and Chen, Meirong and Wu, Bing and Zhang, Hua and Chai, Xiangfei and others},
  journal={Computers in biology and medicine},
  volume={140},
  pages={105063},
  year={2022},
  publisher={Elsevier}
}

@article{creswell2018generative,
  title={Generative adversarial networks: An overview},
  author={Creswell, Antonia and White, Tom and Dumoulin, Vincent and Arulkumaran, Kai and Sengupta, Biswa and Bharath, Anil A},
  journal={IEEE signal processing magazine},
  volume={35},
  number={1},
  pages={53--65},
  year={2018},
  publisher={IEEE}
}

@article{chen2022generative,
  title={Generative adversarial networks in medical image augmentation: A review},
  author={Chen, Yizhou and Yang, Xu-Hua and Wei, Zihan and Heidari, Ali Asghar and Zheng, Nenggan and Li, Zhicheng and Chen, Huiling and Hu, Haigen and Zhou, Qianwei and Guan, Qiu},
  journal={Computers in Biology and Medicine},
  volume={144},
  pages={105382},
  year={2022},
  publisher={Elsevier}
}

@inproceedings{azad2019bi,
  title={Bi-directional ConvLSTM U-Net with densley connected convolutions},
  author={Azad, Reza and Asadi-Aghbolaghi, Maryam and Fathy, Mahmood and Escalera, Sergio},
  booktitle={Proceedings of the IEEE/CVF international conference on computer vision workshops},
  pages={0--0},
  year={2019}
}

@article{shi2015convolutional,
  title={Convolutional LSTM network: A machine learning approach for precipitation nowcasting},
  author={Shi, Xingjian and Chen, Zhourong and Wang, Hao and Yeung, Dit-Yan and Wong, Wai-Kin and Woo, Wang-chun},
  journal={Advances in neural information processing systems},
  volume={28},
  year={2015}
}

@article{zhuang2019evaluation,
  title={Evaluation of algorithms for multi-modality whole heart segmentation: an open-access grand challenge},
  author={Zhuang, Xiahai and Li, Lei and Payer, Christian and {\v{S}}tern, Darko and Urschler, Martin and Heinrich, Mattias P and Oster, Julien and Wang, Chunliang and Smedby, {\"O}rjan and Bian, Cheng and others},
  journal={Medical image analysis},
  volume={58},
  pages={101537},
  year={2019},
  publisher={Elsevier}
}

@article{cetin2023attri,
  title={Attri-VAE: Attribute-based interpretable representations of medical images with variational autoencoders},
  author={Cetin, Irem and Stephens, Maialen and Camara, Oscar and Ballester, Miguel A Gonz{\'a}lez},
  journal={Computerized Medical Imaging and Graphics},
  volume={104},
  pages={102158},
  year={2023},
  publisher={Elsevier}
}

@article{gu2021cyclegan,
  title={CycleGAN denoising of extreme low-dose cardiac CT using wavelet-assisted noise disentanglement},
  author={Gu, Jawook and Yang, Tae Seong and Ye, Jong Chul and Yang, Dong Hyun},
  journal={Medical image analysis},
  volume={74},
  pages={102209},
  year={2021},
  publisher={Elsevier}
}

@article{toth20183d,
  title={3D/2D model-to-image registration by imitation learning for cardiac procedures},
  author={Toth, Daniel and Miao, Shun and Kurzendorfer, Tanja and Rinaldi, Christopher A and Liao, Rui and Mansi, Tommaso and Rhode, Kawal and Mountney, Peter},
  journal={International journal of computer assisted radiology and surgery},
  volume={13},
  pages={1141--1149},
  year={2018},
  publisher={Springer}
}

@article{fukushima2020patient,
  title={Patient Based Bull's Eye Map Display of Coronary Artery and Ventricles From Coronary Computed Tomography Angiography},
  author={Fukushima, Kenji and Matsuo, Yuka and Nagao, Michinobu and Sakai, Akiko and Kihara, Nobuyuki and Onishi, Koji and Sakai, Shuji},
  journal={Journal of computer assisted tomography},
  volume={44},
  number={1},
  pages={26--31},
  year={2020},
  publisher={LWW}
}

@article{lecun2015deep,
  title={Deep learning},
  author={LeCun, Yann and Bengio, Yoshua and Hinton, Geoffrey},
  journal={nature},
  volume={521},
  number={7553},
  pages={436--444},
  year={2015},
  publisher={Nature Publishing Group UK London}
}

@inproceedings{donahue2015long,
  title={Long-term recurrent convolutional networks for visual recognition and description},
  author={Donahue, Jeffrey and Anne Hendricks, Lisa and Guadarrama, Sergio and Rohrbach, Marcus and Venugopalan, Subhashini and Saenko, Kate and Darrell, Trevor},
  booktitle={Proceedings of the IEEE conference on computer vision and pattern recognition},
  pages={2625--2634},
  year={2015}
}

@article{wang2023deep,
  title={RL-I2IT: Image-to-image translation with deep reinforcement learning},
  author={Hu, Jing and Luo, Ziwei and  Feng, Chengming and Hu, Shu and Zhu, Bin and Wu, Xi and Zhu, Hongtu and Li, Xin and Lyu, Siwei and Wang, Xin},
  journal={Neural Networks},
  year={2025}
}

@article{vaswani2017attention,
  title={Attention is all you need},
  author={Vaswani, Ashish and Shazeer, Noam and Parmar, Niki and Uszkoreit, Jakob and Jones, Llion and Gomez, Aidan N and Kaiser, {\L}ukasz and Polosukhin, Illia},
  journal={Advances in neural information processing systems},
  volume={30},
  year={2017}
}

@inproceedings{long2015fully,
  title={Fully convolutional networks for semantic segmentation},
  author={Long, Jonathan and Shelhamer, Evan and Darrell, Trevor},
  booktitle={Proceedings of the IEEE conference on computer vision and pattern recognition},
  pages={3431--3440},
  year={2015}
}

@article{ho2020denoising,
  title={Denoising diffusion probabilistic models},
  author={Ho, Jonathan and Jain, Ajay and Abbeel, Pieter},
  journal={Advances in neural information processing systems},
  volume={33},
  pages={6840--6851},
  year={2020}
}

@article{goodfellow2020generative,
  title={Generative adversarial networks},
  author={Goodfellow, Ian and Pouget-Abadie, Jean and Mirza, Mehdi and Xu, Bing and Warde-Farley, David and Ozair, Sherjil and Courville, Aaron and Bengio, Yoshua},
  journal={Communications of the ACM},
  volume={63},
  number={11},
  pages={139--144},
  year={2020},
  publisher={ACM New York, NY, USA}
}

@article{zhao2023clip,
  title={Clip in medical imaging: A comprehensive survey},
  author={Zhao, Zihao and Liu, Yuxiao and Wu, Han and Wang, Mei and Li, Yonghao and Wang, Sheng and Teng, Lin and Liu, Disheng and Cui, Zhiming and Wang, Qian and others},
  journal={Medical Image Analysis},
  year={2025}
}

@article{yang2023diffusion,
  title={Diffusion models: A comprehensive survey of methods and applications},
  author={Yang, Ling and Zhang, Zhilong and Song, Yang and Hong, Shenda and Xu, Runsheng and Zhao, Yue and Zhang, Wentao and Cui, Bin and Yang, Ming-Hsuan},
  journal={ACM Computing Surveys},
  volume={56},
  number={4},
  pages={1--39},
  year={2023},
  publisher={ACM New York, NY, USA}
}

@article{sutton2018reinforcement,
  title={Reinforcement learning: An introduction},
  author={Sutton, Richard S},
  journal={A Bradford Book},
  year={2018}
}

@inproceedings{tsai2024uu,
  title={UU-Mamba: uncertainty-aware u-mamba for cardiac image segmentation},
  author={Tsai, Ting Yu and Lin, Li and Hu, Shu and Chang, Ming-Ching and Zhu, Hongtu and Wang, Xin},
  booktitle={2024 IEEE 7th International Conference on Multimedia Information Processing and Retrieval (MIPR)},
  pages={267--273},
  year={2024},
  organization={IEEE}
}

@article{christensen2024vision,
  title={Vision--language foundation model for echocardiogram interpretation},
  author={Christensen, Matthew and Vukadinovic, Milos and Yuan, Neal and Ouyang, David},
  journal={Nature Medicine},
  pages={1--8},
  year={2024},
  publisher={Nature Publishing Group US New York}
}

@article{zakeri2023dragnet,
  title={DragNet: Learning-based deformable registration for realistic cardiac MR sequence generation from a single frame},
  author={Zakeri, Arezoo and Hokmabadi, Alireza and Bi, Ning and Wijesinghe, Isuru and Nix, Michael G and Petersen, Steffen E and Frangi, Alejandro F and Taylor, Zeike A and Gooya, Ali},
  journal={Medical Image Analysis},
  volume={83},
  pages={102678},
  year={2023},
  publisher={Elsevier}
}

@article{zakeri2022probabilistic,
  title={A probabilistic deep motion model for unsupervised cardiac shape anomaly assessment},
  author={Zakeri, Arezoo and Hokmabadi, Alireza and Ravikumar, Nishant and Frangi, Alejandro F and Gooya, Ali},
  journal={Medical Image Analysis},
  volume={75},
  pages={102276},
  year={2022},
  publisher={Elsevier}
}

@article{lin2024robust,
  title={Robust covid-19 detection in ct images with clip},
  author={Lin, Li and Krubha, Yamini Sri and Yang, Zhenhuan and Ren, Cheng and Wang, Xin and Hu, Shu},
  journal={IEEE 7th International Conference on Multimedia Information Processing and Retrieval (MIPR)},
  year={2024}
}

@article{olawade2024advancements,
  title={Advancements and applications of Artificial Intelligence in cardiology: Current trends and future prospects},
  author={Olawade, David B and Aderinto, Nicholas and Olatunji, Gbolahan and Kokori, Emmanuel and David-Olawade, Aanuoluwapo C and Hadi, Manizha},
  journal={Journal of Medicine, Surgery, and Public Health},
  pages={100109},
  year={2024},
  publisher={Elsevier}
}

@article{makynen2022wearable,
  title={Wearable Devices Combined with Artificial Intelligence—A Future Technology for Atrial Fibrillation Detection?},
  author={M{\"a}kynen, Marko and Ng, G Andre and Li, Xin and Schlindwein, Fernando S},
  journal={Sensors},
  volume={22},
  number={22},
  pages={8588},
  year={2022},
  publisher={MDPI}
}

@article{tcheandjieu2022large,
  title={Large-scale genome-wide association study of coronary artery disease in genetically diverse populations},
  author={Tcheandjieu, Catherine and Zhu, Xiang and Hilliard, Austin T and Clarke, Shoa L and Napolioni, Valerio and Ma, Shining and Lee, Kyung Min and Fang, Huaying and Chen, Fei and Lu, Yingchang and others},
  journal={Nature medicine},
  volume={28},
  number={8},
  pages={1679--1692},
  year={2022},
  publisher={Nature Publishing Group US New York}
}

@article{christophersen2009familial,
  title={Familial aggregation of atrial fibrillation: a study in Danish twins},
  author={Christophersen, Ingrid Elisabeth and Ravn, Lasse Steen and Budtz-Joergensen, Esben and Skytthe, Axel and Haunsoe, Stig and Svendsen, Jesper Hastrup and Christensen, Kaare},
  journal={Circulation: Arrhythmia and Electrophysiology},
  volume={2},
  number={4},
  pages={378--383},
  year={2009},
  publisher={Am Heart Assoc}
}

@article{buniello2019nhgri,
  title={The NHGRI-EBI GWAS Catalog of published genome-wide association studies, targeted arrays and summary statistics 2019},
  author={Buniello, Annalisa and MacArthur, Jacqueline A L and Cerezo, Maria and Harris, Laura W and Hayhurst, James and Malangone, Cinzia and McMahon, Aoife and Morales, Joannella and Mountjoy, Edward and Sollis, Elliot and others},
  journal={Nucleic acids research},
  volume={47},
  number={D1},
  pages={D1005--D1012},
  year={2019},
  publisher={Oxford University Press}
}

@article{aragam2022discovery,
  title={Discovery and systematic characterization of risk variants and genes for coronary artery disease in over a million participants},
  author={Aragam, Krishna G and Jiang, Tao and Goel, Anuj and Kanoni, Stavroula and Wolford, Brooke N and Atri, Deepak S and Weeks, Elle M and Wang, Minxian and Hindy, George and Zhou, Wei and others},
  journal={Nature Genetics},
  volume={54},
  number={12},
  pages={1803--1815},
  year={2022},
  publisher={Nature Publishing Group US New York}
}

@article{sing2003genes,
  title={Genes, environment, and cardiovascular disease},
  author={Sing, Charles F and Steng{\^a}rd, Jari H and Kardia, Sharon LR},
  journal={Arteriosclerosis, thrombosis, and vascular biology},
  volume={23},
  number={7},
  pages={1190--1196},
  year={2003},
  publisher={Am Heart Assoc}
}

@article{zheng2019explainable,
  title={Explainable cardiac pathology classification on cine MRI with motion characterization by semi-supervised learning of apparent flow},
  author={Zheng, Qiao and Delingette, Herv{\'e} and Ayache, Nicholas},
  journal={Medical image analysis},
  volume={56},
  pages={80--95},
  year={2019},
  publisher={Elsevier}
}

@article{zhang2019deep,
  title={Deep learning for diagnosis of chronic myocardial infarction on nonenhanced cardiac cine MRI},
  author={Zhang, Nan and Yang, Guang and Gao, Zhifan and Xu, Chenchu and Zhang, Yanping and Shi, Rui and Keegan, Jennifer and Xu, Lei and Zhang, Heye and Fan, Zhanming and others},
  journal={Radiology},
  volume={291},
  number={3},
  pages={606--617},
  year={2019},
  publisher={Radiological Society of North America}
}

@article{juffermans20224d,
  title={4D Flow MRI in Ascending Aortic Aneurysms: Reproducibility of Hemodynamic Parameters},
  author={Juffermans, Joe F and van Assen, Hans C and te Kiefte, Bastiaan JC and Ramaekers, Mitch JFG and van der Palen, Roel LF and van den Boogaard, Pieter and Adriaans, Bouke P and Wildberger, Joachim E and Dekkers, Ilona A and Scholte, Arthur JHA and others},
  journal={Applied Sciences},
  volume={12},
  number={8},
  pages={3912},
  year={2022},
  publisher={MDPI}
}

@article{zhuang2021role,
  title={The role of 4D flow MRI for clinical applications in cardiovascular disease: current status and future perspectives},
  author={Zhuang, Baiyan and Sirajuddin, Arlene and Zhao, Shihua and Lu, Minjie},
  journal={Quantitative Imaging in Medicine and Surgery},
  volume={11},
  number={9},
  pages={4193},
  year={2021},
  publisher={AME Publications}
}

@article{kiricsli2013standardized,
  title={Standardized evaluation framework for evaluating coronary artery stenosis detection, stenosis quantification and lumen segmentation algorithms in computed tomography angiography},
  author={Kiri{\c{s}}li, HA and Schaap, Michiel and Metz, CT and Dharampal, AS and Meijboom, Willem Bob and Papadopoulou, Stella-Lida and Dedic, Admir and Nieman, Koen and de Graaf, Michiel A and Meijs, MFL and others},
  journal={Medical image analysis},
  volume={17},
  number={8},
  pages={859--876},
  year={2013},
  publisher={Elsevier}
}

@article{rinaldi2022invasive,
  title={Invasive Functional Coronary Assessment in Myocardial Ischemia with Non-Obstructive Coronary Arteries: From Pathophysiological Mechanisms to Clinical Implications},
  author={Rinaldi, Riccardo and Salzillo, Carmine and Caff{\`e}, Andrea and Montone, Rocco A},
  journal={Reviews in Cardiovascular Medicine},
  volume={23},
  number={11},
  pages={371},
  year={2022},
  publisher={IMR Press}
}

@article{barratt2023environmental,
  title={Environmental impact of cardiovascular healthcare},
  author={Barratt, Alexandra L and Li, Yan and Gooroovadoo, Isabelle and Todd, Allyson and Dou, Yuanlong and McAlister, Scott and Semsarian, Christopher},
  journal={Open Heart},
  volume={10},
  number={1},
  pages={e002279},
  year={2023},
  publisher={Archives of Disease in childhood}
}

@article{bhatnagar2017environmental,
  title={Environmental determinants of cardiovascular disease},
  author={Bhatnagar, Aruni},
  journal={Circulation research},
  volume={121},
  number={2},
  pages={162--180},
  year={2017},
  publisher={Am Heart Assoc}
}

@article{powell2022social,
  title={Social determinants of cardiovascular disease},
  author={Powell-Wiley, Tiffany M and Baumer, Yvonne and Baah, Foster Osei and Baez, Andrew S and Farmer, Nicole and Mahlobo, Christa T and Pita, Mario A and Potharaju, Kameswari A and Tamura, Kosuke and Wallen, Gwenyth R},
  journal={Circulation research},
  volume={130},
  number={5},
  pages={782--799},
  year={2022},
  publisher={Am Heart Assoc}
}

@article{munzel2022environmental,
  title={Environmental risk factors and cardiovascular diseases: a comprehensive expert review},
  author={M{\"u}nzel, Thomas and Hahad, Omar and others},
  journal={Cardiovascular Research},
  year={2022},
  publisher={Oxford University Press US}
}

@article{jafari2023automated,
  title={Automated diagnosis of cardiovascular diseases from cardiac magnetic resonance imaging using deep learning models: A review},
  author={Jafari, Mahboobeh and Shoeibi, Afshin and Khodatars, Marjane and Ghassemi, Navid and Moridian, Parisa and Alizadehsani, Roohallah and Khosravi, Abbas and Ling, Sai Ho and Delfan, Niloufar and Zhang, Yu-Dong and others},
  journal={Computers in Biology and Medicine},
  pages={106998},
  year={2023},
  publisher={Elsevier}
}

@article{michelhaugh2023using,
  title={Using artificial intelligence to better predict and develop biomarkers},
  author={Michelhaugh, Sam A and Januzzi, James L},
  journal={Clinics in Laboratory Medicine},
  volume={43},
  number={1},
  pages={99--114},
  year={2023},
  publisher={Elsevier}
}

@article{vasan2006biomarkers,
  title={Biomarkers of cardiovascular disease: molecular basis and practical considerations},
  author={Vasan, Ramachandran S},
  journal={Circulation},
  year={2006},
  publisher={Am Heart Assoc}
}

@article{jone2022artificial,
  title={Artificial Intelligence in Congenital Heart Disease: Current State and Prospects},
  author={Jone, Pei-Ni and Gearhart, Addison and Lei, Howard and Xing, Fuyong and Nahar, Jai and Lopez-Jimenez, Francisco and Diller, Gerhard-Paul and Marelli, Ariane and Wilson, Laura and Saidi, Arwa and others},
  journal={JACC: Advances},
  volume={1},
  number={5},
  pages={100153},
  year={2022},
  publisher={American College of Cardiology Foundation Washington DC}
}

@article{lecun2015deepronneberger2015u,
  title={Deep learning},
  author={LeCun, Yann and Bengio, Yoshua and Hinton, Geoffrey},
  journal={nature},
  volume={521},
  number={7553},
  pages={436--444},
  year={2015},
  publisher={Nature Publishing Group UK London}
}

@article{chandrashekar2021deep,
  title={A Deep Learning Approach to Visualize Aortic Aneurysm Morphology Without the Use of Intravenous Contrast Agents},
  author={Chandrashekar, Anirudh and Handa, Ashok and Lapolla, Pierfrancesco and Shivakumar, Natesh and Uberoi, Raman and Grau, Vicente and Lee, Regent},
  journal={Annals of Surgery},
  year={2021},
  publisher={Wolters Kluwer Health}
}

@article{guo2022artificial,
  title={Artificial intelligence study on left ventricular function among normal individuals, hypertrophic cardiomyopathy and dilated cardiomyopathy patients using 1.5 T cardiac cine MR images obtained by SSFP sequence},
  author={Guo, Jiajun and Lu, HongFei and Chen, Yinyin and Zeng, Mengsu and Jin, Hang},
  journal={The British Journal of Radiology},
  volume={95},
  number={1133},
  pages={20201060},
  year={2022},
  publisher={The British Institute of Radiology.}
}

@article{kusunose2021standardize,
  title={How to standardize the measurement of left ventricular ejection fraction},
  author={Kusunose, Kenya and Zheng, Robert and Yamada, Hirotsugu and Sata, Masataka},
  journal={Journal of Medical Ultrasonics},
  pages={1--9},
  year={2021},
  publisher={Springer}
}

@article{zhou2021artificial,
  title={Artificial intelligence in echocardiography: detection, functional evaluation, and disease diagnosis},
  author={Zhou, Jia and Du, Meng and Chang, Shuai and Chen, Zhiyi},
  journal={Cardiovascular ultrasound},
  volume={19},
  number={1},
  pages={1--11},
  year={2021},
  publisher={BioMed Central}
}

@article{akerman2023automated,
  title={Automated echocardiographic detection of heart failure with preserved ejection fraction using artificial intelligence},
  author={Akerman, Ashley P and Porumb, Mihaela and Scott, Christopher G and Beqiri, Arian and Chartsias, Agisilaos and Ryu, Alexander J and Hawkes, William and Huntley, Geoffrey D and Arystan, Ayana Z and Kane, Garvan C and others},
  journal={JACC: Advances},
  volume={2},
  number={6},
  pages={100452},
  year={2023},
  publisher={Elsevier}
}

@article{he2023blinded,
  title={Blinded, randomized trial of sonographer versus AI cardiac function assessment},
  author={He, Bryan and Kwan, Alan C and Cho, Jae Hyung and Yuan, Neal and Pollick, Charles and Shiota, Takahiro and Ebinger, Joseph and Bello, Natalie A and Wei, Janet and Josan, Kiranbir and others},
  journal={Nature},
  volume={616},
  number={7957},
  pages={520--524},
  year={2023},
  publisher={Nature Publishing Group UK London}
}

@article{bizopoulos2018deep,
  title={Deep learning in cardiology},
  author={Bizopoulos, Paschalis and Koutsouris, Dimitrios},
  journal={IEEE reviews in biomedical engineering},
  volume={12},
  pages={168--193},
  year={2018},
  publisher={IEEE}
}

@article{liu2023deep,
  title={A deep learning framework assisted echocardiography with diagnosis, lesion localization, phenogrouping heterogeneous disease, and anomaly detection},
  author={Liu, Bohan and Chang, Hao and Yang, Dong and Yang, Feifei and Wang, Qiushuang and Deng, Yujiao and Li, Lijun and Lv, Wenqing and Zhang, Bo and Yu, Liheng and others},
  journal={Scientific Reports},
  volume={13},
  number={1},
  pages={3},
  year={2023},
  publisher={Nature Publishing Group UK London}
}

@article{krittanawong2023deep,
  title={Deep Learning for Echocardiography: Introduction for Clinicians and Future Vision: State-of-the-Art Review},
  author={Krittanawong, Chayakrit and others},
  journal={Life},
  year={2023},
  publisher={MDPI}
}

@article{bueno2023qrs,
  title={QRS-T angles as markers for heart sphericity in subjects with intrauterine growth restriction: a simulation study},
  author={Bueno-Palomeque, Freddy L and Mountris, Konstantinos A and Ortigosa, Nuria and Bail{\'o}n, Raquel and Bijnens, Bart and Crispi, F{\`a}tima and Pueyo, Esther and Minchol{\'e}, Ana and Laguna, Pablo},
  journal={IEEE Journal of Biomedical and Health Informatics},
  year={2023},
  publisher={IEEE}
}

@article{de2023framing,
  title={The framing of time-dependent machine learning models improves risk estimation among young individuals with acute coronary syndromes},
  author={de Carvalho, Luiz S{\'e}rgio Fernandes and Alexim, Gustavo and Nogueira, Ana Claudia Cavalcante and Fernandez, Marta Duran and Rezende, Tito Barbosa and Avila, Sandra and Reis, Ricardo Torres Bispo and Soares, Alexandre Anderson Munhoz and Sposito, Andrei Carvalho},
  journal={Scientific Reports},
  volume={13},
  number={1},
  pages={1021},
  year={2023},
  publisher={Nature Publishing Group UK London}
}

@article{d2023deep,
  title={Deep learning to detect significant coronary artery disease from plain chest radiographs AI4CAD},
  author={D'Ancona, Giuseppe and Massussi, Mauro and Savardi, Mattia and Signoroni, Alberto and Di Bacco, Lorenzo and Farina, Davide and Metra, Marco and Maroldi, Roberto and Muneretto, Claudio and Ince, H{\"u}seyin and others},
  journal={International Journal of Cardiology},
  volume={370},
  pages={435--441},
  year={2023},
  publisher={Elsevier}
}

@article{smirnova2023spontaneous,
  title={Spontaneous coronary artery dissection: an unpredictable event},
  author={Smirnova, Alexandra and Aliberti, Flaminia and Cavaliere, Claudia and Gatti, Ilaria and Vilardo, Viviana and Giorgianni, Carmelina and Cassani, Chiara and Repetto, Alessandra and Narula, Nupoor and Giuliani, Lorenzo and others},
  journal={European Heart Journal Supplements},
  volume={25},
  number={Supplement\_B},
  pages={B7--B11},
  year={2023},
  publisher={Oxford University Press US}
}

@article{krittanawong2021machine,
  title={Machine learning and deep learning to predict mortality in patients with spontaneous coronary artery dissection},
  author={Krittanawong, Chayakrit and Virk, Hafeez Ul Hassan and Kumar, Anirudh and Aydar, Mehmet and Wang, Zhen and Stewart, Matthew P and Halperin, Jonathan L},
  journal={Scientific reports},
  volume={11},
  number={1},
  pages={8992},
  year={2021},
  publisher={Nature Publishing Group UK London}
}

@inproceedings{mokhtari2022echognn,
  title={EchoGNN: Explainable Ejection Fraction Estimation with Graph Neural Networks},
  author={Mokhtari, Masoud and Tsang, Teresa and Abolmaesumi, Purang and Liao, Renjie},
  booktitle={International Conference on Medical Image Computing and Computer-Assisted Intervention},
  pages={360--369},
  year={2022},
  organization={Springer}
}

@article{litjens2019state,
  title={State-of-the-art deep learning in cardiovascular image analysis},
  author={Litjens, Geert and Ciompi, Francesco and Wolterink, Jelmer M and de Vos, Bob D and Leiner, Tim and Teuwen, Jonas and I{\v{s}}gum, Ivana},
  journal={JACC: Cardiovascular imaging},
  volume={12},
  number={8 Part 1},
  pages={1549--1565},
  year={2019},
  publisher={American College of Cardiology Foundation Washington, DC}
}

@article{doudesis2023machine,
  title={Machine learning for diagnosis of myocardial infarction using cardiac troponin concentrations},
  author={Doudesis, Dimitrios and Lee, Kuan Ken and Boeddinghaus, Jasper and Bularga, Anda and Ferry, Amy V and Tuck, Chris and Lowry, Matthew TH and Lopez-Ayala, Pedro and Nestelberger, Thomas and Koechlin, Luca and others},
  journal={Nature Medicine},
  pages={1--10},
  year={2023},
  publisher={Nature Publishing Group US New York}
}

@article{cho2020artificial,
  title={Artificial intelligence algorithm for detecting myocardial infarction using six-lead electrocardiography},
  author={Cho, Younghoon and Kwon, Joon-myoung and Kim, Kyung-Hee and Medina-Inojosa, Jose R and Jeon, Ki-Hyun and Cho, Soohyun and Lee, Soo Youn and Park, Jinsik and Oh, Byung-Hee},
  journal={Scientific reports},
  volume={10},
  number={1},
  pages={20495},
  year={2020},
  publisher={Nature Publishing Group UK London}
}

@article{madan2022hybrid,
  title={A hybrid deep learning approach for ECG-based arrhythmia classification},
  author={Madan, Parul and Singh, Vijay and Singh, Devesh Pratap and Diwakar, Manoj and Pant, Bhaskar and Kishor, Avadh},
  journal={Bioengineering},
  volume={9},
  number={4},
  pages={152},
  year={2022},
  publisher={MDPI}
}

@article{kolk2023machine,
  title={Machine learning of electrophysiological signals for the prediction of ventricular arrhythmias: systematic review and examination of heterogeneity between studies},
  author={Kolk, Maarten ZH and Deb, Brototo and Ruip{\'e}rez-Campillo, Samuel and Bhatia, Neil K and Clopton, Paul and Wilde, Arthur AM and Narayan, Sanjiv M and Knops, Reinoud E and Tjong, Fleur VY},
  journal={EBioMedicine},
  volume={89},
  year={2023},
  publisher={Elsevier}
}

@article{pantelidis2023deep,
  title={Deep learning to diagnose left ventricular hypertrophy from standard, 12-lead ECG signals: a proof-of-concept study},
  author={Pantelidis, P and Oikonomou, E and Souvaliotis, N and Spartalis, M and Lampsas, S and Bampa, M and Bakogiannis, C and Antonopoulos, A and Siasos, G and Vavuranakis, M and others},
  journal={Europace},
  volume={25},
  number={Supplement\_1},
  pages={euad122--534},
  year={2023},
  publisher={Oxford University Press US}
}

@article{liu2021deep,
  title={Deep learning-based automated left ventricular ejection fraction assessment using 2-D echocardiography},
  author={Liu, Xin and Fan, Yiting and Li, Shuang and Chen, Meixiang and Li, Ming and Hau, William Kongto and Zhang, Heye and Xu, Lin and Lee, Alex Pui-Wai},
  journal={American Journal of Physiology-Heart and Circulatory Physiology},
  volume={321},
  number={2},
  pages={H390--H399},
  year={2021},
  publisher={American Physiological Society Rockville, MD}
}

@article{yang2023application,
  title={Application of Artificial Intelligence-Based Auxiliary Diagnosis in Congenital Heart Disease Screening.},
  author={Yang, Hongbo and Pan, Jiahua and Wang, Weilian and Guo, Tao and Ma, Tengyuan},
  journal={Anatolian Journal of Cardiology/Anadolu Kardiyoloji Dergisi},
  volume={27},
  number={4},
  year={2023}
}

@article{zhao2022deep,
  title={Deep learning assessment of left ventricular hypertrophy based on electrocardiogram},
  author={Zhao, Xiaoli and Huang, Guifang and Wu, Lin and Wang, Min and He, Xuemin and Wang, Jyun-Rong and Zhou, Bin and Liu, Yong and Lin, Yesheng and Liu, Dinghui and others},
  journal={Frontiers in Cardiovascular Medicine},
  volume={9},
  pages={952089},
  year={2022},
  publisher={Frontiers}
}

@article{hsu2022machine,
  title={Machine learning for electrocardiographic features to identify left atrial enlargement in young adults: CHIEF Heart study},
  author={Hsu, Chu-Yu and Liu, Pang-Yen and Liu, Shu-Hsin and Kwon, Younghoon and Lavie, Carl J and Lin, Gen-Min},
  journal={Frontiers in Cardiovascular Medicine},
  volume={9},
  pages={840585},
  year={2022},
  publisher={Frontiers}
}

@article{ter2023juvenile,
  title={Juvenile-onset multifocal atrial arrhythmias, atrial standstill and compound heterozygosity of genetic variants in TAF1A: sentinel event for evolving dilated cardiomyopathy--a case report},
  author={ter Bekke, Rachel MA and de Schouwer, Koen and Conti, Sergio and Claes, Godelieve RF and Vanoevelen, Jo and Gommers, Suzanne and Helderman-van den Enden, Apollonia TJM and Brunner-LaRocca, Hans-Peter},
  journal={European Heart Journal-Case Reports},
  pages={ytad255},
  year={2023},
  publisher={Oxford University Press}
}

@article{wang2022deep,
  title={Deep-Learning-Based Detection of Paroxysmal Supraventricular Tachycardia Using Sinus-Rhythm Electrocardiograms},
  author={Wang, Lei and Dang, Shipeng and Chen, Shuangxiong and Sun, Jin-Yu and Wang, Ru-Xing and Pan, Feng},
  journal={Journal of Clinical Medicine},
  volume={11},
  number={15},
  pages={4578},
  year={2022},
  publisher={MDPI}
}

@article{raicea2021giant,
  title={Giant left atrial myxoma--literature review and case presentation},
  author={Raicea, Victor Cornel and Suciu, Hora{\c{t}}iu and Raicea, Andrei Dan and Macarie, Gheorghe Cosmin and Mezei, Tibor and Maier, Maria Smaranda},
  journal={Romanian Journal of Morphology and Embryology},
  volume={62},
  number={2},
  pages={361},
  year={2021},
  publisher={Romanian Academy Publishing House}
}

@article{niimi2022machine,
  title={Machine learning models for prediction of adverse events after percutaneous coronary intervention},
  author={Niimi, Nozomi and Shiraishi, Yasuyuki and Sawano, Mitsuaki and Ikemura, Nobuhiro and Inohara, Taku and Ueda, Ikuko and Fukuda, Keiichi and Kohsaka, Shun},
  journal={Scientific reports},
  volume={12},
  number={1},
  pages={6262},
  year={2022},
  publisher={Nature Publishing Group UK London}
}

@article{crabb2023deep,
  title={Deep Learning Subtraction Angiography: Improved Generalizability with Transfer Learning},
  author={Crabb, Brendan T and Hamrick, Forrest and Richards, Tyler and Eiswirth, Preston and Noo, Frederic and Hsiao, Albert and Fine, Gabriel C},
  journal={Journal of Vascular and Interventional Radiology},
  volume={34},
  number={3},
  pages={409--419},
  year={2023},
  publisher={Elsevier}
}

@article{apostolopoulos2023deep,
  title={Deep learning-enhanced nuclear medicine SPECT imaging applied to cardiac studies},
  author={Apostolopoulos, Ioannis D and Papandrianos, Nikolaos I and Feleki, Anna and Moustakidis, Serafeim and Papageorgiou, Elpiniki I},
  journal={EJNMMI physics},
  volume={10},
  number={1},
  pages={6},
  year={2023},
  publisher={Springer}
}

@article{alskaf2022deep,
  title={Deep learning applications in myocardial perfusion imaging, a systematic review and meta-analysis},
  author={Alskaf, Ebraham and Dutta, Utkarsh and Scannell, Cian M and Chiribiri, Amedeo},
  journal={Informatics in Medicine Unlocked},
  pages={101055},
  year={2022},
  publisher={Elsevier}
}

@article{ghorbani2020deep,
  title={Deep learning interpretation of echocardiograms},
  author={Ghorbani, Amirata and Ouyang, David and Abid, Abubakar and He, Bryan and Chen, Jonathan H and Harrington, Robert A and Liang, David H and Ashley, Euan A and Zou, James Y},
  journal={NPJ digital medicine},
  volume={3},
  number={1},
  pages={10},
  year={2020},
  publisher={Nature Publishing Group UK London}
}

@article{eng2021automated,
  title={Automated coronary calcium scoring using deep learning with multicenter external validation},
  author={Eng, David and Chute, Christopher and Khandwala, Nishith and Rajpurkar, Pranav and Long, Jin and Shleifer, Sam and Khalaf, Mohamed H and Sandhu, Alexander T and Rodriguez, Fatima and Maron, David J and others},
  journal={NPJ digital medicine},
  volume={4},
  number={1},
  pages={88},
  year={2021},
  publisher={Nature Publishing Group UK London}
}

@article{bruns2020deep,
  title={Deep learning from dual-energy information for whole-heart segmentation in dual-energy and single-energy non-contrast-enhanced cardiac CT},
  author={Bruns, Steffen and Wolterink, Jelmer M and Takx, Richard AP and van Hamersvelt, Robbert W and Such{\'a}, Dominika and Viergever, Max A and Leiner, Tim and I{\v{s}}gum, Ivana},
  journal={Medical physics},
  volume={47},
  number={10},
  pages={5048--5060},
  year={2020},
  publisher={Wiley Online Library}
}

@article{peper2020functional,
  title={Functional cardiac CT--going beyond anatomical evaluation of coronary artery disease with Cine CT, CT-FFR, CT perfusion and machine learning},
  author={Peper, Joyce and Such{\'a}, Dominika and Swaans, Martin and Leiner, Tim},
  journal={The British Journal of Radiology},
  volume={93},
  number={1113},
  pages={20200349},
  year={2020},
  publisher={The British Institute of Radiology.}
}

@article{lin2022deep,
  title={Deep learning-enabled coronary CT angiography for plaque and stenosis quantification and cardiac risk prediction: an international multicentre study},
  author={Lin, Andrew and Manral, Nipun and McElhinney, Priscilla and Killekar, Aditya and Matsumoto, Hidenari and Kwiecinski, Jacek and Pieszko, Konrad and Razipour, Aryabod and Grodecki, Kajetan and Park, Caroline and others},
  journal={The Lancet Digital Health},
  volume={4},
  number={4},
  pages={e256--e265},
  year={2022},
  publisher={Elsevier}
}

@incollection{wang2024u,
  title={U-medsam: Uncertainty-aware medsam for medical image segmentation},
  author={Wang, Xin and Liu, Xiaoyu and Huang, Peng and Huang, Pu and Hu, Shu and Zhu, Hongtu},
  booktitle={Medical Image Segmentation Challenge},
  pages={206--217},
  year={2024},
  publisher={Springer}
}

@article{xu2023coronary,
  title={Coronary artery stent evaluation by CTA: impact of deep learning reconstruction and subtraction technique},
  author={Xu, Cheng and Yi, Yan and Xu, Min and Yan, Jing and Guo, Yu-Bo and Wang, Jian and Wang, Yun and Li, Yu-Mei and Jin, Zheng-Yu and Wang, Yi-Ning},
  journal={American Journal of Roentgenology},
  volume={220},
  number={1},
  pages={63--72},
  year={2023},
  publisher={Am Roentgen Ray Soc}
}

@article{garg2023role,
  title={Role of Deep Learning in Computed Tomography},
  author={Garg, Yash and Seetharam, Karthik and Sharma, Manjari and Rohita, Dipesh K and Nabi, Waseem},
  journal={Cureus},
  volume={15},
  number={5},
  year={2023},
  publisher={Cureus}
}

@article{kustner2020cinenet,
  title={CINENet: deep learning-based 3D cardiac CINE MRI reconstruction with multi-coil complex-valued 4D spatio-temporal convolutions},
  author={K{\"u}stner, Thomas and Fuin, Niccolo and Hammernik, Kerstin and Bustin, Aurelien and Qi, Haikun and Hajhosseiny, Reza and Masci, Pier Giorgio and Neji, Radhouene and Rueckert, Daniel and Botnar, Ren{\'e} M and others},
  journal={Scientific reports},
  volume={10},
  number={1},
  pages={13710},
  year={2020},
  publisher={Nature Publishing Group UK London}
}

@article{wahab2023developing,
  title={Developing a Deep-Learning-Based Coronary Artery Disease Detection Technique Using Computer Tomography Images},
  author={Wahab Sait, Abdul Rahaman and Dutta, Ashit Kumar},
  journal={Diagnostics},
  volume={13},
  number={7},
  pages={1312},
  year={2023},
  publisher={MDPI}
}

@article{lionakis2022spontaneous,
  title={Spontaneous coronary artery dissection: A review of diagnostic methods and management strategies},
  author={Lionakis, Nikolaos and Briasoulis, Alexandros and Zouganeli, Virginia and Dimopoulos, Stavros and Kalpakos, Dionisios and Kourek, Christos},
  journal={World Journal of Cardiology},
  volume={14},
  number={10},
  pages={522},
  year={2022},
  publisher={Baishideng Publishing Group Inc}
}

@article{gao2020learning,
  title={Learning physical properties in complex visual scenes: An intelligent machine for perceiving blood flow dynamics from static CT angiography imaging},
  author={Gao, Zhifan and Wang, Xin and others},
  journal={Neural Networks},
  year={2020},
  publisher={Elsevier}
}

@article{catapano20204d,
  title={4D flow imaging of the thoracic aorta: is there an added clinical value?},
  author={Catapano, Federica and Pambianchi, Giacomo and Cundari, Giulia and Rebelo, Jo{\~a}o and Cilia, Francesco and Carbone, Iacopo and Catalano, Carlo and Francone, Marco and Galea, Nicola},
  journal={Cardiovascular Diagnosis and Therapy},
  volume={10},
  number={4},
  pages={1068},
  year={2020},
  publisher={AME Publications}
}

@article{iannattone2020artificial,
  title={Artificial intelligence for diagnosis of acute coronary syndromes: a meta-analysis of machine learning approaches},
  author={Iannattone, Patrick A and Zhao, Xun and VanHouten, Jacob and Garg, Akhil and Huynh, Thao},
  journal={Canadian Journal of Cardiology},
  volume={36},
  number={4},
  pages={577--583},
  year={2020},
  publisher={Elsevier}
}

@article{zhang2022artificial,
  title={Artificial intelligence in cardiovascular atherosclerosis imaging},
  author={Zhang, Jia and Han, Ruijuan and Shao, Guo and Lv, Bin and Sun, Kai},
  journal={Journal of Personalized Medicine},
  volume={12},
  number={3},
  pages={420},
  year={2022},
  publisher={MDPI}
}

@article{awad2014public,
  title={Public knowledge of cardiovascular disease and its risk factors in Kuwait: a cross-sectional survey},
  author={Awad, Abdelmoneim and Al-Nafisi, Hala},
  journal={BMC public health},
  volume={14},
  number={1},
  pages={1--11},
  year={2014},
  publisher={BioMed Central}
}

@article{mezquita2023clinical,
  title={Clinical quantitative coronary artery stenosis and coronary atherosclerosis imaging: a Consensus Statement from the Quantitative Cardiovascular Imaging Study Group},
  author={M{\'e}zquita, Aldo J V{\'a}zquez and others},
  journal={Nature Reviews Cardiology},
  year={2023}
}

@article{ramaekers2023clinician,
  title={A clinician’s guide to understanding aortic 4D flow MRI},
  author={Ramaekers, Mitch JFG and Westenberg, Jos JM and Adriaans, Bouke P and Nijssen, Estelle C and Wildberger, Joachim E and Lamb, Hildo J and Schalla, Simon},
  journal={Insights into Imaging},
  volume={14},
  number={1},
  pages={1--14},
  year={2023},
  publisher={SpringerOpen}
}

@article{itu2016machine,
  title={A machine-learning approach for computation of fractional flow reserve from coronary computed tomography},
  author={Itu, Lucian and Rapaka, Saikiran and Passerini, Tiziano and Georgescu, Bogdan and Schwemmer, Chris and Schoebinger, Max and Flohr, Thomas and Sharma, Puneet and Comaniciu, Dorin},
  journal={Journal of applied physiology},
  volume={121},
  number={1},
  pages={42--52},
  year={2016},
  publisher={American Physiological Society Bethesda, MD}
}

@article{kong2020learning,
  title={Learning tree-structured representation for 3D coronary artery segmentation},
  author={Kong, Bin and Wang, Xin and Bai, Junjie and Lu, Yi and Gao, Feng and Cao, Kunlin and Xia, Jun and Song, Qi and Yin, Youbing},
  journal={Computerized Medical Imaging and Graphics},
  volume={80},
  pages={101688},
  year={2020},
  publisher={Elsevier}
}

@article{berry2014fractional,
  title={Fractional flow reserve, coronary flow reserve and the index of microvascular resistance in clinical practice},
  author={Berry, Colin and others},
  journal={Radcliffe Cardiology. Feb},
  pages={1--6},
  year={2014}
}

@article{gao2022clinical,
  title={A clinical trial for computed tomography myocardial perfusion based non-invasive index of microcirculatory resistance (MPBIMR): Rationale and trial design},
  author={Gao, Beibei and Zhu, Darong and Xie, Jianchang and Wu, Bokai and Xu, Peng and Liu, Jia and Tong, Xiaoshan and Chen, Rongliang and Zhu, Lijun and Zhou, Liang and others},
  journal={American Journal of Translational Research},
  volume={14},
  number={8},
  pages={5552},
  year={2022},
  publisher={e-Century Publishing Corporation}
}

@article{leclerc2019deep,
  title={Deep learning for segmentation using an open large-scale dataset in 2D echocardiography},
  author={Leclerc, Sarah and Smistad, Erik and Pedrosa, Joao and {\O}stvik, Andreas and Cervenansky, Frederic and Espinosa, Florian and Espeland, Torvald and Berg, Erik Andreas Rye and Jodoin, Pierre-Marc and Grenier, Thomas and others},
  journal={IEEE transactions on medical imaging},
  volume={38},
  number={9},
  pages={2198--2210},
  year={2019},
  publisher={IEEE}
}

@article{bissell20234d,
  title={4D Flow cardiovascular magnetic resonance consensus statement: 2023 update},
  author={Bissell, Malenka M and Raimondi, Francesca and Ait Ali, Lamia and Allen, Bradley D and Barker, Alex J and Bolger, Ann and Burris, Nicholas and Carh{\"a}ll, Carl-Johan and Collins, Jeremy D and Ebbers, Tino and others},
  journal={Journal of Cardiovascular Magnetic Resonance},
  volume={25},
  number={1},
  pages={1--24},
  year={2023},
  publisher={BioMed Central}
}

@article{saeed2015cardiac,
  title={Cardiac MR imaging: current status and future direction},
  author={Saeed, Maythem and Van, Tu Anh and Krug, Roland and Hetts, Steven W and Wilson, Mark W},
  journal={Cardiovascular diagnosis and therapy},
  volume={5},
  number={4},
  pages={290},
  year={2015},
  publisher={AME Publications}
}

@article{li2022medical,
  title={Medical image analysis on left atrial LGE MRI for atrial fibrillation studies: A review},
  author={Li, Lei and Zimmer, Veronika A and Schnabel, Julia A and Zhuang, Xiahai},
  journal={Medical image analysis},
  pages={102360},
  year={2022},
  publisher={Elsevier}
}

@article{sun2023social,
  title={Social determinants, cardiovascular disease, and health care cost: a nationwide study in the United States using machine learning},
  author={Sun, Feinuo and Yao, Jie and Du, Shichao and others},
  journal={Journal of the American Heart Association},
  year={2023}
}

@article{li2020atrial,
  title={Atrial scar quantification via multi-scale CNN in the graph-cuts framework},
  author={Li, Lei and Wu, Fuping and Yang, Guang and Xu, Lingchao and Wong, Tom and Mohiaddin, Raad and Firmin, David and Keegan, Jennifer and Zhuang, Xiahai},
  journal={Medical image analysis},
  volume={60},
  pages={101595},
  year={2020},
  publisher={Elsevier}
}

@article{zreik2018recurrent,
  title={A recurrent CNN for automatic detection and classification of coronary artery plaque and stenosis in coronary CT angiography},
  author={Zreik, Majd and Van Hamersvelt, Robbert W and Wolterink, Jelmer M and Leiner, Tim and Viergever, Max A and I{\v{s}}gum, Ivana},
  journal={IEEE transactions on medical imaging},
  volume={38},
  number={7},
  pages={1588--1598},
  year={2018},
  publisher={IEEE}
}

@article{ferdian20204dflownet,
  title={4DFlowNet: super-resolution 4D flow MRI using deep learning and computational fluid dynamics},
  author={Ferdian, Edward and Suinesiaputra, Avan and Dubowitz, David J and Zhao, Debbie and Wang, Alan and Cowan, Brett and Young, Alistair A},
  journal={Frontiers in Physics},
  pages={138},
  year={2020},
  publisher={Frontiers}
}

@article{zeng2022imagecas,
  title={ImageCAS: A Large-Scale Dataset and Benchmark for Coronary Artery Segmentation based on Computed Tomography Angiography Images},
  author={Zeng, An and Wu, Chunbiao and Huang, Meiping and Zhuang, Jian and Bi, Shanshan and Pan, Dan and Ullah, Najeeb and Khan, Kaleem Nawaz and Wang, Tianchen and Shi, Yiyu and others},
  journal={Computerized Medical Imaging and Graphics},
  year={2023}
}

@article{yan2022impact,
  title={Impact of Pressure Wire on Fractional Flow Reserve and Hemodynamics of the Coronary Arteries: A Computational and Clinical Study},
  author={Yan, Zhengzheng and Yao, Zhifeng and Guo, Weifeng and Shang, Dandan and Chen, Rongliang and Liu, Jia and Cai, Xiao-Chuan and Ge, Junbo},
  journal={IEEE Transactions on Biomedical Engineering},
  volume={70},
  number={5},
  pages={1683--1691},
  year={2022},
  publisher={IEEE}
}

@article{raghunath2021deep,
  title={Deep neural networks can predict new-onset atrial fibrillation from the 12-lead ECG and help identify those at risk of atrial fibrillation--related stroke},
  author={Raghunath, Sushravya and Pfeifer, John M and Ulloa-Cerna, Alvaro E and Nemani, Arun and Carbonati, Tanner and Jing, Linyuan and vanMaanen, David P and Hartzel, Dustin N and Ruhl, Jeffery A and Lagerman, Braxton F and others},
  journal={Circulation},
  volume={143},
  number={13},
  pages={1287--1298},
  year={2021},
  publisher={Am Heart Assoc}
}

@article{ren2022comparison,
  title={Comparison of a deep learning-accelerated T2-weighted turbo spin echo sequence and its conventional counterpart for female pelvic MRI: reduced acquisition times and improved image quality},
  author={Ren, Jing and Li, Yuan and Liu, Fei-Shi and Liu, Chong and Zhu, Jin-Xia and Nickel, Marcel Dominik and Wang, Xiao-Ye and Liu, Xin-Yu and Zhao, Jia and He, Yong-Lan and others},
  journal={Insights into Imaging},
  volume={13},
  number={1},
  pages={193},
  year={2022},
  publisher={Springer}
}

@article{ueda2022deep,
  title={Deep learning reconstruction of diffusion-weighted MRI improves image quality for prostatic imaging},
  author={Ueda, Takahiro and Ohno, Yoshiharu and Yamamoto, Kaori and Murayama, Kazuhiro and Ikedo, Masato and Yui, Masao and Hanamatsu, Satomu and Tanaka, Yumi and Obama, Yuki and Ikeda, Hirotaka and others},
  journal={Radiology},
  volume={303},
  number={2},
  pages={373--381},
  year={2022},
  publisher={Radiological Society of North America}
}

@article{berhane2022deep,
  title={Deep learning--based velocity antialiasing of 4D-flow MRI},
  author={Berhane, Haben and Scott, Michael B and Barker, Alex J and McCarthy, Patrick and Avery, Ryan and Allen, Brad and Malaisrie, Chris and Robinson, Joshua D and Rigsby, Cynthia K and Markl, Michael},
  journal={Magnetic resonance in medicine},
  volume={88},
  number={1},
  pages={449--463},
  year={2022},
  publisher={Wiley Online Library}
}

@article{xue2020automated,
  title={Automated inline analysis of myocardial perfusion MRI with deep learning},
  author={Xue, Hui and Davies, Rhodri H and Brown, Louise AE and Knott, Kristopher D and Kotecha, Tushar and Fontana, Marianna and Plein, Sven and Moon, James C and Kellman, Peter},
  journal={Radiology: Artificial Intelligence},
  volume={2},
  number={6},
  pages={e200009},
  year={2020},
  publisher={Radiological Society of North America}
}

@article{jiang2020detection,
  title={Detection of left atrial enlargement using a convolutional neural network-enabled electrocardiogram},
  author={Jiang, Junrong and Deng, Hai and Xue, Yumei and Liao, Hongtao and Wu, Shulin},
  journal={Frontiers in Cardiovascular Medicine},
  volume={7},
  pages={609976},
  year={2020},
  publisher={Frontiers Media SA}
}

@article{matsuoka2022deep,
  title={Deep learning-based approach for detecting signs of atrial septal defect on chest radiographs: a proof of concept study},
  author={Matsuoka, Ryo and Akazawa, Hiroshi and Kodera, Satoshi and Soma, Katsura and Yagi, Hiroki and Umei, Masahiko and Kadowaki, Hiroshi and Ishida, Junichi and Shinohara, Hiroki and Katsushika, Susumu and others},
  journal={medRxiv},
  pages={2022--01},
  year={2022},
  publisher={Cold Spring Harbor Laboratory Press}
}

@inproceedings{ye2021deeptag,
  title={Deeptag: An unsupervised deep learning method for motion tracking on cardiac tagging magnetic resonance images},
  author={Ye, Meng and Kanski, Mikael and Yang, Dong and Chang, Qi and Yan, Zhennan and Huang, Qiaoying and Axel, Leon and Metaxas, Dimitris},
  booktitle={Proceedings of the IEEE/CVF conference on computer vision and pattern recognition},
  pages={7261--7271},
  year={2021}
}

@article{chen2020deep,
  title={Deep learning for cardiac image segmentation: a review},
  author={Chen, Chen and Qin, Chen and Qiu, Huaqi and Tarroni, Giacomo and Duan, Jinming and Bai, Wenjia and Rueckert, Daniel},
  journal={Frontiers in Cardiovascular Medicine},
  volume={7},
  year={2020},
  publisher={Frontiers Media SA}
}

@article{li2023multi,
  title={Multi-modality cardiac image computing: A survey},
  author={Li, Lei and Ding, Wangbin and Huang, Liqun and Zhuang, Xiahai and Grau, Vicente},
  journal={Medical Image Analysis},
  year={2023},
  publisher={Elsevier}
}

@inproceedings{yang2020cpr,
  title={Cpr-gcn: Conditional partial-residual graph convolutional network in automated anatomical labeling of coronary arteries},
  author={Yang, Han and Zhen, Xingjian and Chi, Ying and Zhang, Lei and Hua, Xian-Sheng},
  booktitle={IEEE CVPR},
  year={2020}
}

@article{zhao2023heart,
  title={Heart-brain connections: Phenotypic and genetic insights from magnetic resonance images},
  author={Zhao, Bingxin and Li, Tengfei and Fan, Zirui and Yang, Yue and Shu, Juan and Yang, Xiaochen and Wang, Xifeng and Luo, Tianyou and Tang, Jiarui and Xiong, Di and others},
  journal={Science},
  year={2023},
  publisher={American Association for the Advancement of Science}
}

@article{wu2019automated,
  title={Automated anatomical labeling of coronary arteries via bidirectional tree LSTMs},
  author={Wu, Dan and Wang, Xin and Bai, Junjie and Xu, Xiaoyang and Ouyang, Bin and Li, Yuwei and Zhang, Heye and Song, Qi and Cao, Kunlin and Yin, Youbing},
  journal={International journal of computer assisted radiology and surgery},
  publisher={Springer}
}

@article{fu2021review,
  title={A review of deep learning based methods for medical image multi-organ segmentation},
  author={Fu, Yabo and Lei, Yang and Wang, Tonghe and Curran, Walter J and Liu, Tian and Yang, Xiaofeng},
  journal={Physica Medica},
  year={2021},
  publisher={Elsevier}
}

@article{zhang2023challenges,
  title={On the Challenges and Perspectives of Foundation Models for Medical Image Analysis},
  author={Zhang, Shaoting and Metaxas, Dimitris},
  journal={Medical Image Analysis},
  year={2024}
}

@article{ma2019iterative,
  title={An iterative multi-path fully convolutional neural network for automatic cardiac segmentation in cine MR images},
  author={Ma, Zongqing and Wu, Xi and Wang, Xin and Song, Qi and Yin, Youbing and Cao, Kunlin and Wang, Yan and Zhou, Jiliu},
  journal={Medical Physics},
  year={2019},
  publisher={Wiley Online Library}
}

@article{lu2023attention,
  title={Attention-driven tree-structured convolutional LSTM for high dimensional data understanding},
  author={Lu, Yi and Kong, Bin and Gao, Feng and Cao, Kunlin and Lyu, Siwei and Zhang, Shaoting and Hu, Shu and Yin, Youbing and Wang, Xin},
  journal={Frontiers in Physics},
  year={2023},
  publisher={Frontiers}
}

@article{zhou2023comprehensive,
  title={A comprehensive survey on pretrained foundation models: A history from bert to chatgpt},
  author={Zhou, Ce and Li, Qian and Li, Chen and Yu, Jun and Liu, Yixin and Wang, Guangjing and Zhang, Kai and Ji, Cheng and Yan, Qiben and He, Lifang and others},
  journal={International Journal of Machine Learning and Cybernetics},
  year={2024}
}

@article{radford2018improving,
  title={Improving language understanding by generative pre-training},
  author={Radford, Alec and Narasimhan, Karthik and Salimans, Tim and Sutskever, Ilya and others},
  year={2018},
  publisher={OpenAI}
}

@article{nguyen2023lvm,
  title={Lvm-med: Learning large-scale self-supervised vision models for medical imaging via second-order graph matching},
  author={MH Nguyen, Duy and Nguyen, Hoang and Diep, Nghiem and Pham, Tan Ngoc and Cao, Tri and Nguyen, Binh and Swoboda, Paul and Ho, Nhat and Albarqouni, Shadi and Xie, Pengtao and others},
  journal={Advances in Neural Information Processing Systems},
  volume={36},
  pages={27922--27950},
  year={2023}
}

@article{sun2023survey,
  title={A Survey of Reasoning with Foundation Models},
  author={Sun, Jiankai and Zheng, Chuanyang and Xie, Enze and Liu, Zhengying and Chu, Ruihang and Qiu, Jianing and Xu, Jiaqi and Ding, Mingyu and Li, Hongyang and Geng, Mengzhe and others},
  journal={ACM Computing Surveys},
  year={2025}
}

@article{wang2023medfmc,
  title={A real-world dataset and benchmark for foundation model adaptation in medical image classification},
  author={Wang, Dequan and Wang, Xiaosong and Wang, Lilong and Li, Mengzhang and Da, Qian and Liu, Xiaoqiang and Gao, Xiangyu and Shen, Jun and He, Junjun and Shen, Tian and others},
  journal={Scientific Data},
  volume={10},
  number={1},
  pages={574},
  year={2023},
  publisher={Nature Publishing Group UK London}
}

@inproceedings{wang2023foundation,
  title={Foundation model for endoscopy video analysis via large-scale self-supervised pre-train},
  author={Wang, Zhao and Liu, Chang and Zhang, Shaoting and Dou, Qi},
  booktitle={International Conference on Medical Image Computing and Computer-Assisted Intervention},
  year={2023},
  organization={Springer}
}

@article{zhou2023foundation,
  title={A foundation model for generalizable disease detection from retinal images},
  author={Zhou, Yukun and Chia, Mark A and Wagner, Siegfried K and Ayhan, Murat S and Williamson, Dominic J and Struyven, Robbert R and Liu, Timing and Xu, Moucheng and Lozano, Mateo G and Woodward-Court, Peter and others},
  journal={Nature},
  year={2023},
  publisher={Nature Publishing Group UK London}
}

@inproceedings{mildenhall2021nerf,
  title={Nerf: Representing scenes as neural radiance fields for view synthesis},
  author={Mildenhall, Ben and Srinivasan, Pratul P and Tancik, Matthew and Barron, Jonathan T and Ramamoorthi, Ravi and Ng, Ren},
  journal={Communications of the ACM},
  volume={65},
  number={1},
  year={2021},
  publisher={ACM New York, NY, USA}
}

@inproceedings{corona2022mednerf,
  title={Mednerf: Medical neural radiance fields for reconstructing 3d-aware ct-projections from a single x-ray},
  author={Corona-Figueroa, Abril and Frawley, Jonathan and others},
  booktitle={EMBC},
  year={2022},
  organization={IEEE}
}

@article{lo2012extraction,
  title={Extraction of airways from CT (EXACT'09)},
  author={Lo, Pechin and Van Ginneken, Bram and Reinhardt, Joseph M and others},
  journal={IEEE Transactions on Medical Imaging},
  volume={31},
  number={11},
  year={2012},
  publisher={IEEE}
}

@article{maken20232d,
  title={2D-to-3D: A Review for Computational 3D Image Reconstruction from x-ray Images},
  author={Maken, Payal and Gupta, Abhishek},
  journal={Archives of Computational Methods in Engineering},
  year={2023}
}

@article{hu2023umednerf,
  title={UMedNeRF: Uncertainty-aware Single View Volumetric Rendering for Medical Neural Radiance Fields},
  author={Hu, Jing and Fan, Qinrui and Hu, Shu and Lyu, Siwei and Wu, Xi and Wang, Xin},
  journal={IEEE International Symposium on Biomedical Imaging},
  year={2024}
}

@article{holste2023severe,
  title={Severe aortic stenosis detection by deep learning applied to echocardiography},
  author={Holste, Gregory and Oikonomou, Evangelos K and Mortazavi, Bobak J and Coppi, Andreas and Faridi, Kamil F and Miller, Edward J and Forrest, John K and McNamara, Robert L and Ohno-Machado, Lucila and Yuan, Neal and others},
  journal={European Heart Journal},
  volume={44},
  number={43},
  pages={4592--4604},
  year={2023},
  publisher={Oxford University Press US}
}

@article{li2023multimodal,
  title={Multimodal foundation models: From specialists to general-purpose assistants},
  author={Li, Chunyuan and Gan, Zhe and Yang, Zhengyuan and Yang, Jianwei and Li, Linjie and Wang, Lijuan and Gao, Jianfeng and others},
  journal={Foundations and Trends{\textregistered} in Computer Graphics and Vision},
  volume={16},
  number={1-2},
  pages={1--214},
  year={2024},
  publisher={Now Publishers, Inc.}
}

@article{li2023llava,
  title={Llava-med: Training a large language-and-vision assistant for biomedicine in one day},
  author={Li, Chunyuan and Wong, Cliff and Zhang, Sheng and Usuyama, Naoto and Liu, Haotian and Yang, Jianwei and Naumann, Tristan and Poon, Hoifung and Gao, Jianfeng},
  journal={Advances in Neural Information Processing Systems},
  volume={36},
  pages={28541--28564},
  year={2023}
}

@article{bello2019deep,
  title={Deep-learning cardiac motion analysis for human survival prediction},
  author={Bello, Ghalib A and Dawes, Timothy JW and Duan, Jinming and Biffi, Carlo and De Marvao, Antonio and Howard, Luke SGE and Gibbs, J Simon R and Wilkins, Martin R and Cook, Stuart A and Rueckert, Daniel and others},
  journal={Nature machine intelligence},
  year={2019},
  publisher={Nature Publishing Group UK London}
}

@inproceedings{mascarenhas2021comparison,
  title={A comparison between VGG16, VGG19 and ResNet50 architecture frameworks for Image Classification},
  author={Mascarenhas, Sheldon and Agarwal, Mukul},
  booktitle={IEEE CENTCON},
  year={2021}
}

@article{dhawan2004role,
  title={Role of magnetic resonance imaging in visualizing coronary arteries},
  author={Dhawan, Sumeesh and Dharmashankar, Kodlipet C and Tak, Tahir},
  journal={Clinical medicine \& research},
  year={2004},
  publisher={Marshfield Clinic}
}

@article{li2001magnetic,
  title={Magnetic resonance imaging of coronary arteries},
  author={Li, Debiao and Deshpande, Vibhas},
  journal={Topics in Magnetic Resonance Imaging},
  volume={12},
  number={5},
  year={2001},
  publisher={LWW}
}

@article{botnar2000noninvasive,
  title={Noninvasive coronary vessel wall and plaque imaging with magnetic resonance imaging},
  author={Botnar, Ren{\'e} M and Stuber, Matthias and Kissinger, Kraig V and Kim, Won Y and Spuentrup, Elmar and Manning, Warren J},
  journal={Circulation},
  year={2000},
  publisher={Am Heart Assoc}
}

@article{van1999magnetic,
  title={Magnetic resonance imaging of the coronary arteries: clinical results from three dimensional evaluation of a respiratory gated technique},
  author={Van Geuns, RJM and De Bruin, HG and Rensing, BJWM and Wielopolski, PA and Hulshoff, MD and Van Ooijen, PMA and Oudkerk, Matthijs and De Feyter, PJ},
  journal={Heart},
  year={1999},
  publisher={BMJ Publishing Group Ltd and British Cardiovascular Society}
}

@article{pennell1996assessment,
  title={Assessment of coronary artery stenosis by magnetic resonance imaging.},
  author={Pennell, Dudley J and Bogren, Hugo G and Keegan, Jennifer and Firmin, David N and Underwood, S Richard},
  journal={Heart},
  year={1996},
  publisher={BMJ Publishing Group}
}

@article{van1995magnetic,
  title={Magnetic resonance imaging in coronary artery disease},
  author={van der Wall, Ernst E and Vliegen, Hubert W and de Roos, Albert and Bruschke, Albert VG},
  journal={Circulation},
  year={1995},
  publisher={Am Heart Assoc}
}

@article{pennell1993magnetic,
  title={Magnetic resonance imaging of coronary arteries: technique and preliminary results.},
  author={Pennell, DJ and Keegan, J and Firmin, DN and Gatehouse, PD and Underwood, SR and Longmore, DB},
  journal={British heart journal},
  year={1993},
  publisher={BMJ Publishing Group}
}

@article{peng2016review,
  title={A review of heart chamber segmentation for structural and functional analysis using cardiac magnetic resonance imaging},
  author={Peng, Peng and Lekadir, Karim and Gooya, Ali and Shao, Ling and Petersen, Steffen E and Frangi, Alejandro F},
  journal={Magnetic Resonance Materials in Physics, Biology and Medicine},
  year={2016},
  publisher={Springer}
}

@incollection{ehrhardt2022autoencoders,
  title={Autoencoders and variational autoencoders in medical image analysis},
  author={Ehrhardt, Jan and Wilms, Matthias},
  booktitle={Biomedical Image Synthesis and Simulation},
  year={2022},
  publisher={Elsevier}
}

@article{update2017heart,
  title={Heart disease and stroke statistics--2017 update},
  author={Update, AHA Statistical},
  journal={Circulation},
  volume={135},
  pages={e146--e603},
  year={2017},
  publisher={Am Heart Assoc}
}

@article{de2002intravascular,
  title={Intravascular ultrasound elastography: an overview},
  author={De Korte, Chris L and Van Der Steen, Anton FW},
  journal={Ultrasonics},
  year={2002},
  publisher={Elsevier}
}

@article{xu2020fundamentals,
  title={Fundamentals and role of intravascular ultrasound in percutaneous coronary intervention},
  author={Xu, James and Lo, Sidney},
  journal={Cardiovascular Diagnosis and Therapy},
  year={2020},
  publisher={AME Publications}
}

@article{aly2021cardiac,
  title={Cardiac ultrasound: an anatomical and clinical review},
  author={Aly, Islam and Rizvi, Asad and others},
  journal={Translational Research in Anatomy},
  year={2021},
  publisher={Elsevier}
}

@article{li2020survey,
  title={A survey of heart anomaly detection using ambulatory electrocardiogram (ECG)},
  author={Li, Hongzu and Boulanger, Pierre},
  journal={Sensors},
  year={2020},
  publisher={MDPI}
}

@article{yu2018coronary,
  title={Coronary plaque characteristics on baseline CT predict the need for late revascularization in symptomatic patients after percutaneous intervention},
  author={Yu, Mengmeng and Lu, Zhigang and Li, Wenbin and Wei, Meng and Yan, Jing and Zhang, Jiayin},
  journal={European Radiology},
  year={2018},
  publisher={Springer}
}

@article{slomka2021quantitative,
  title={Quantitative clinical nuclear cardiology, part 2: Evolving/emerging applications},
  author={Slomka, Piotr J and Moody, Jonathan B and Miller, Robert JH and Renaud, Jennifer M and Ficaro, Edward P and Garcia, Ernest V},
  journal={Journal of Nuclear Cardiology},
  year={2021},
  publisher={Springer}
}

@article{lin2023coronary,
  title={Coronary heart disease prediction method fusing domain-adaptive transfer learning with graph convolutional networks (GCN)},
  author={Lin, Huizhong and Chen, Kaizhi and Xue, Yutao and Zhong, Shangping and Chen, Lianglong and Ye, Mingfang},
  journal={Scientific Reports},
  volume={13},
  number={1},
  pages={14276},
  year={2023},
  publisher={Nature Publishing Group UK London}
}

@article{shenoy2024novel,
  title={A Novel 3D Camera-based ECG-Imaging System for Electrode Position Discovery and Heart-Torso Registration},
  author={Shenoy, Nikhil and Toloubidokhti, Maryam and Gharbia, Omar and Khoshknab, Mirmilad P and Nazarian, Saman and Sapp, John L and Singh, Vivek and Kapoor, Ankur and Wang, Linwei},
  journal={Authorea Preprints},
  year={2024},
  publisher={Authorea}
}

@article{fu2023prior,
  title={Prior skeleton based online deep reinforcement learning for coronary artery centerline extraction},
  author={Fu, Zeyu and Fu, Zhuang and Fang, Zi and Wang, Zehao and Fei, Jian and Xie, Rongli and Han, Hui},
  journal={Proceedings of the Institution of Mechanical Engineers, Part H: Journal of Engineering in Medicine},
  year={2023},
  publisher={SAGE Publications Sage UK: London, England}
}

@article{murat2021review,
  title={Review of deep learning-based atrial fibrillation detection studies},
  author={Murat, Fatma and Sadak, Ferhat and Yildirim, Ozal and Talo, Muhammed and Murat, Ender and Karabatak, Murat and Demir, Yakup and Tan, Ru-San and Acharya, U Rajendra},
  journal={International journal of environmental research and public health},
  year={2021}
}

@article{kuchynka2015role,
  title={The role of magnetic resonance imaging and cardiac computed tomography in the assessment of left atrial anatomy, size, and function},
  author={Kuchynka, Petr and Podzimkova, Jana and Masek, Martin and Lambert, Lukas and Cerny, Vladimir and Danek, Barbara and Palecek, Tomas and others},
  journal={BioMed research international},
  volume={2015},
  year={2015},
  publisher={Hindawi}
}

@article{vukadinovic2023deep,
  title={Deep learning-enabled analysis of medical images identifies cardiac sphericity as an early marker of cardiomyopathy and related outcomes},
  author={Vukadinovic, Milos and Kwan, Alan C and Yuan, Victoria and Salerno, Michael and Lee, Daniel C and Albert, Christine M and Cheng, Susan and Li, Debiao and Ouyang, David and Clarke, Shoa L},
  journal={Med},
  year={2023},
  publisher={Elsevier}
}

@article{wang2023deephcd,
  title={Deep learning for discrimination of hypertrophic cardiomyopathy and hypertensive heart disease on MRI native T1 maps},
  author={Wang, Zi-Chen and Fan, Zhang-Zhengyi and Liu, Xi-Yuan and Zhu, Ming-Jie and Jiang, Shan-Shan and Tian, Song and Chen, Bing-Hua and Wu, Lian-Ming},
  journal={Journal of Magnetic Resonance Imaging},
  year={2023},
  publisher={Wiley Online Library}
}

@inproceedings{gautam2022current,
  title={Current and future applications of artificial intelligence in coronary artery disease},
  author={Gautam, Nitesh and Saluja, Prachi and Malkawi, Abdallah and Rabbat, Mark G and Al-Mallah, Mouaz H and Pontone, Gianluca and Zhang, Yiye and Lee, Benjamin C and Al’Aref, Subhi J},
  booktitle={Healthcare},
  year={2022},
  organization={MDPI}
}

@article{zhao2023agmn,
  title={AGMN: Association graph-based graph matching network for coronary artery semantic labeling on invasive coronary angiograms},
  author={Zhao, Chen and Xu, Zhihui and Jiang, Jingfeng and Esposito, Michele and Pienta, Drew and Hung, Guang-Uei and Zhou, Weihua},
  journal={Pattern recognition},
  volume={143},
  pages={109789},
  year={2023},
  publisher={Elsevier}
}

@inproceedings{li2020segmentation,
  title={Segmentation to label: Automatic coronary artery labeling from mask parcellation},
  author={Li, Zhuowei and Xia, Qing and Wang, Wenji and Yan, Zhennan and Yin, Ruohan and Pan, Changjie and Metaxas, Dimitris},
  booktitle={Machine Learning in Medical Imaging},
  year={2020},
  organization={Springer}
}

@article{papamanolis2021myocardial,
  title={Myocardial perfusion simulation for coronary artery disease: A coupled patient-specific multiscale model},
  author={Papamanolis, Lazaros and Kim, Hyun Jin and Jaquet, Clara and Sinclair, Matthew and Schaap, Michiel and Danad, Ibrahim and van Diemen, Pepijn and Knaapen, Paul and Najman, Laurent and Talbot, Hugues and others},
  journal={Annals of biomedical engineering},
  year={2021},
  publisher={Springer}
}

@article{fawaz2023invasive,
  title={Invasive Detection of Coronary Microvascular Dysfunction: How It Began, and Where We Are Now},
  author={Fawaz, Samer and Khan, Sarosh and Simpson, Rupert and Clesham, Gerald and Cook, Christopher M and Davies, John R and Karamasis, Grigoris V and Keeble, Thomas R},
  journal={Interventional Cardiology: Reviews, Research, Resources},
  year={2023},
  publisher={Radcliffe Cardiology}
}

@article{teng2022survey,
  title={A survey on the interpretability of deep learning in medical diagnosis},
  author={Teng, Qiaoying and Liu, Zhe and Song, Yuqing and Han, Kai and Lu, Yang},
  journal={Multimedia Systems},
  year={2022},
  publisher={Springer}
}

@article{coorey2022health,
  title={The health digital twin to tackle cardiovascular disease—a review of an emerging interdisciplinary field},
  author={Coorey, Genevieve and Figtree, Gemma A and Fletcher, David F and Snelson, Victoria J and Vernon, Stephen Thomas and Winlaw, David and Grieve, Stuart M and McEwan, Alistair and Yang, Jean Yee Hwa and Qian, Pierre and others},
  journal={NPJ digital medicine},
  year={2022},
  publisher={Nature Publishing Group UK London}
}

@article{corral2020digital,
  title={The ‘Digital Twin’to enable the vision of precision cardiology},
  author={Corral-Acero, Jorge and Margara, Francesca and Marciniak, Maciej and Rodero, Cristobal and Loncaric, Filip and Feng, Yingjing and Gilbert, Andrew and Fernandes, Joao F and Bukhari, Hassaan A and Wajdan, Ali and others},
  journal={European heart journal},
  year={2020},
  publisher={Oxford University Press}
}

@article{ghatti2023digital,
  title={Digital Twins in Healthcare: A Survey of Current Methods},
  author={Ghatti, Siddharth and Yurish, Livvy Ann and Shen, Haiying and Rheuban, Karen and Enfield, Kyle B and Facteau, Nikki Reyer and Engel, Gina and Dowdell, Kim},
  journal={Archives of Clinical and Biomedical Research},
  year={2023},
  publisher={Fortune Journals}
}

@article{pirola20194,
  title={4-D flow MRI-based computational analysis of blood flow in patient-specific aortic dissection},
  author={Pirola, Selene and Guo, Baolei and Menichini, Claudia and Saitta, Simone and Fu, Weiguo and Dong, Zhihui and Xu, Xiao Yun},
  journal={IEEE Transactions on Biomedical Engineering},
  year={2019},
  publisher={IEEE}
}

@article{serrano2023coronary,
  title={Coronary artery segmentation based on transfer learning and UNet architecture on computed tomography coronary angiography images},
  author={Serrano-Ant{\'o}n, Bel{\'e}n and Otero-Cacho, Alberto and L{\'o}pez-Otero, Diego and D{\'\i}az-Fern{\'a}ndez, Brais and Bastos-Fern{\'a}ndez, Mar{\'\i}a and P{\'e}rez-Mu{\~n}uzuri, Vicente and Gonz{\'a}lez-Juanatey, Jos{\'e} Ram{\'o}n and Mu{\~n}uzuri, Alberto P},
  journal={IEEE Access},
  year={2023},
  publisher={IEEE}
}

@inproceedings{kolli2018image,
  title={Image-Based Computational Fluid Dynamic Analysis for Surgical Planning of Sequential Grafts in Coronary Artery Bypass Grafting},
  author={Kolli, Kranthi K and Min, James K},
  booktitle={2018 40th Annual International Conference of the IEEE Engineering in Medicine and Biology Society (EMBC)},
  year={2018},
  organization={IEEE}
}

@article{baessato2020incremental,
  title={The incremental role of coronary computed tomography in chronic coronary syndromes},
  author={Baessato, Francesca and Guglielmo, Marco and Muscogiuri, Giuseppe and Baggiano, Andrea and Fusini, Laura and Scafuri, Stefano and Babbaro, Mario and Mollace, Rocco and Collevecchio, Ada and Guaricci, Andrea I and others},
  journal={Journal of Clinical Medicine},
  year={2020},
  publisher={MDPI}
}

@article{isgum2012automatic,
  title={Automatic coronary calcium scoring in low-dose chest computed tomography},
  author={Isgum, Ivana and Prokop, Mathias and Niemeijer, Meindert and Viergever, Max A and Van Ginneken, Bram},
  journal={IEEE transactions on medical imaging},
  year={2012},
  publisher={IEEE}
}

@article{bui2020simultaneous,
  title={Simultaneous multi-structure segmentation of the heart and peripheral tissues in contrast enhanced cardiac computed tomography angiography},
  author={Bui, Vy and Shanbhag, Sujata M and Levine, Oscar and Jacobs, Matthew and Bandettini, W Patricia and Chang, Lin-Ching and Chen, Marcus Y and Hsu, Li-Yueh},
  journal={IEEE Access},
  year={2020},
  publisher={IEEE}
}

@inproceedings{liu2021dual,
  title={Dual-cycle constrained bijective vae-gan for tagged-to-cine magnetic resonance image synthesis},
  author={Liu, Xiaofeng and Xing, Fangxu and Prince, Jerry L and Carass, Aaron and Stone, Maureen and El Fakhri, Georges and Woo, Jonghye},
  booktitle={ISBI},
  year={2021},
  organization={IEEE}
}

@inproceedings{jaen2022synthetic,
  title={Synthetic Generation of Cardiac MR Images Combining Convolutional Variational Autoencoders and Style Transfer},
  author={Ja{\'e}n-Lorites, Jos{\'e} Manuel and P{\'e}rez-Pelegr{\'\i}, Manuel and others},
  booktitle={IEEE EMBC},
  year={2022},
  organization={IEEE}
}

@article{kweon2021deep,
  title={Deep reinforcement learning for guidewire navigation in coronary artery phantom},
  author={Kweon, Jihoon and Kim, Kyunghwan and Lee, Chaehyuk and Kwon, Hwi and Park, Jinwoo and Song, Kyoseok and Kim, Young In and Park, Jeeone and Back, Inwook and Roh, Jae-Hyung and others},
  journal={IEEE Access},
  year={2021},
  publisher={IEEE}
}

@inproceedings{trentin2015automatic,
  title={An automatic tool for thoracic aorta segmentation and 3D geometric analysis},
  author={Trentin, Chiara and Faggiano, Elena and Conti, Michele and Auricchio, Ferdinando},
  booktitle={2015 9th International Symposium on Image and Signal Processing and Analysis (ISPA)},
  year={2015},
  organization={IEEE}
}

@article{righini2023four,
  title={Four-Dimensional Flow MRI for the Evaluation of Aortic Endovascular Graft: A Pilot Study},
  author={Righini, Paolo and Secchi, Francesco and Mazzaccaro, Daniela and Giese, Daniel and Galligani, Marina and Avishay, Dor and Capra, Davide and Monti, Caterina Beatrice and Nano, Giovanni},
  journal={Diagnostics},
  year={2023},
  publisher={MDPI}
}

@article{ayub2020coronary,
  title={Coronary microvascular dysfunction and the role of noninvasive cardiovascular imaging},
  author={Ayub, Muhammad Talha and Kalra, Dinesh},
  journal={Diagnostics},
  year={2020},
  publisher={MDPI}
}

@article{vesal2020fully,
  title={Fully automated 3d cardiac mri localisation and segmentation using deep neural networks},
  author={Vesal, Sulaiman and Maier, Andreas and Ravikumar, Nishant},
  journal={Journal of Imaging},
  year={2020},
  publisher={MDPI}
}

@article{clarke2020invasive,
  title={Invasive evaluation of the microvasculature in acute myocardial infarction: coronary flow reserve versus the index of microcirculatory resistance},
  author={Clarke, John-Ross D and Kennedy, Randol and Lau, Freddy Duarte and Lancaster, Gilead I and Zarich, Stuart W},
  journal={Journal of Clinical Medicine},
  volume={9},
  number={1},
  year={2020},
  publisher={Multidisciplinary Digital Publishing Institute (MDPI)}
}

@article{jacobs2021automated,
  title={Automated segmental analysis of fully quantitative myocardial blood flow maps by first-pass perfusion cardiovascular magnetic resonance},
  author={Jacobs, Matthew and Benovoy, Mitchel and Chang, Lin-Ching and Corcoran, David and Berry, Colin and Arai, Andrew E and Hsu, Li-Yueh},
  journal={IEEE access},
  year={2021},
  publisher={IEEE}
}

@article{meng2022mulvimotion,
  title={MulViMotion: Shape-aware 3D Myocardial Motion Tracking from Multi-View Cardiac MRI},
  author={Meng, Qingjie and Qin, Chen and Bai, Wenjia and Liu, Tianrui and de Marvao, Antonio and O’Regan, Declan P and Rueckert, Daniel},
  journal={IEEE transactions on medical imaging},
  year={2022},
  publisher={IEEE}
}

@inproceedings{li2023automated,
  title={Automated Coronary Arteries Labeling Via Geometric Deep Learning},
  author={Li, Yadan and Armin, Mohammad Ali and Denman, Simon and Ahmedt-Aristizabal, David},
  booktitle={2023 IEEE 20th International Symposium on Biomedical Imaging (ISBI)},
  year={2023},
  organization={IEEE}
}

@article{zhao2023explainability,
  title={Explainability for large language models: A survey},
  author={Zhao, Haiyan and Chen, Hanjie and Yang, Fan and Liu, Ninghao and Deng, Huiqi and Cai, Hengyi and Wang, Shuaiqiang and Yin, Dawei and Du, Mengnan},
  journal={ACM Transactions on Intelligent Systems and Technology},
  year={2023},
  publisher={ACM New York, NY}
}

@inproceedings{he2023dmcvr,
  title={DMCVR: Morphology-Guided Diffusion Model for 3D Cardiac Volume Reconstruction},
  author={He, Xiaoxiao and Tan, Chaowei and Han, Ligong and Liu, Bo and Axel, Leon and Li, Kang and Metaxas, Dimitris N},
  booktitle={International Conference on Medical Image Computing and Computer-Assisted Intervention},
  year={2023},
  organization={Springer}
}

@article{kazerouni2023diffusion,
  title={Diffusion models in medical imaging: A comprehensive survey},
  author={Kazerouni, Amirhossein and Aghdam, Ehsan Khodapanah and Heidari, Moein and Azad, Reza and Fayyaz, Mohsen and Hacihaliloglu, Ilker and Merhof, Dorit},
  journal={Medical Image Analysis},
  year={2023},
  publisher={Elsevier}
}

@inproceedings{kim2023diffusion,
title={Diffusion Adversarial Representation Learning for Self-supervised Vessel Segmentation},
author={Boah Kim and Yujin Oh and Jong Chul Ye},
booktitle={The Eleventh International Conference on Learning Representations },
year={2023},
url={https://openreview.net/forum?id=H0gdPxSwkPb}
}

@article{lu2023ultrafast,
  title={Ultrafast Cardiac Imaging Using Deep Learning For Speckle-Tracking Echocardiography},
  author={Lu, Jingfeng and others},
  journal={IEEE TRANSACTIONS ON ULTRASONICS, FERROELECTRICS, AND FREQUENCY CONTROL},
  year={2023}
}

@article{xu2020contrast,
  title={Contrast agent-free synthesis and segmentation of ischemic heart disease images using progressive sequential causal GANs},
  author={Xu, Chenchu and Xu, Lei and others},
  journal={Medical image analysis},
  year={2020},
  publisher={Elsevier}
}

@inproceedings{zheng2011machine,
  title={Machine learning based vesselness measurement for coronary artery segmentation in cardiac CT volumes},
  author={Zheng, Yefeng and Loziczonek, Maciej and Georgescu, Bogdan and Zhou, S Kevin and Vega-Higuera, Fernando and Comaniciu, Dorin},
  booktitle={Medical Imaging 2011: Image Processing},
  volume={7962},
  year={2011},
  organization={Spie}
}

@article{shin2016deep,
  title={Deep convolutional neural networks for computer-aided detection: CNN architectures, dataset characteristics and transfer learning},
  author={Shin, Hoo-Chang and Roth, Holger R and Gao, Mingchen and Lu, Le and others},
  journal={IEEE transactions on medical imaging},
  year={2016},
  publisher={IEEE}
}

@article{ibanez20182017,
  title={2017 ESC Guidelines for the management of acute myocardial infarction in patients presenting with ST-segment elevation: The Task Force for the management of acute myocardial infarction in patients presenting with ST-segment elevation of the European Society of Cardiology (ESC)},
  author={Ibanez, Borja and others},
  journal={European heart journal},
  year={2018},
  publisher={Oxford University Press}
}

@article{tao2018digital,
  title={Digital twin in industry: State-of-the-art},
  author={Tao, Fei and Zhang, He and Liu, Ang and Nee, Andrew YC},
  journal={IEEE Transactions on industrial informatics},
  volume={15},
  number={4},
  year={2018},
  publisher={IEEE}
}

@article{ma2024segment,
  title={Segment anything in medical images},
  author={Ma, Jun and He, Yuting and Li, Feifei and Han, Lin and You, Chenyu and Wang, Bo},
  journal={Nature Communications},
  year={2024},
  publisher={Nature Publishing Group UK London}
}

@article{campello2021multi,
  title={Multi-centre, multi-vendor and multi-disease cardiac segmentation: the M\&Ms challenge},
  author={Campello, Victor M and Gkontra, Polyxeni and Izquierdo, Cristian and Martin-Isla, Carlos and Sojoudi, Alireza and Full, Peter M and Maier-Hein, Klaus and Zhang, Yao and He, Zhiqiang and Ma, Jun and others},
  journal={IEEE Transactions on Medical Imaging},
  year={2021},
  publisher={IEEE}
}

@article{berkaya2018survey,
  title={A survey on ECG analysis},
  author={Berkaya, Selcan Kaplan and Uysal, Alper Kursat and Gunal, Efnan Sora and Ergin, Semih and Gunal, Serkan and Gulmezoglu, M Bilginer},
  journal={Biomedical Signal Processing and Control},
  year={2018},
  publisher={Elsevier}
}

@article{ccimen2016reconstruction,
  title={Reconstruction of coronary arteries from X-ray angiography: A review},
  author={{\c{C}}imen, Serkan and Gooya, Ali and Grass, Michael and Frangi, Alejandro F},
  journal={Medical image analysis},
  year={2016},
  publisher={Elsevier}
}

@article{tamaki2024current,
  title={Current status and perspectives of nuclear cardiology},
  author={Tamaki, Nagara and Manabe, Osamu},
  journal={Annals of Nuclear Medicine},
  year={2024},
  publisher={Springer}
}

@article{alizadehsani2021coronary,
  title={Coronary artery disease detection using artificial intelligence techniques: A survey of trends, geographical differences and diagnostic features 1991--2020},
  author={Alizadehsani, Roohallah and Khosravi, Abbas and Roshanzamir, Mohamad and Abdar, Moloud and Sarrafzadegan, Nizal and Shafie, Davood and Khozeimeh, Fahime and Shoeibi, Afshin and Nahavandi, Saeid and Panahiazar, Maryam and others},
  journal={Computers in Biology and Medicine},
  volume={128},
  pages={104095},
  year={2021},
  publisher={Elsevier}
}

@article{zhao2023early,
  title={Early detection of coronary microvascular dysfunction using machine learning algorithm based on vectorcardiography and cardiodynamicsgram features},
  author={Zhao, Xiaoye and Gong, Yinglan and Zhang, Jucheng and Liu, Haipeng and Huang, Tianhai and Jiang, Jun and Niu, Yanli and Xia, Ling and Mao, Jiandong},
  journal={IRBM},
  volume={44},
  number={6},
  pages={100805},
  year={2023},
  publisher={Elsevier}
}

@article{wang2022current,
  title={Current advancement in diagnosing atrial fibrillation by utilizing wearable devices and artificial intelligence: A review study},
  author={Wang, Yu-Chiang and Xu, Xiaobo and Hajra, Adrija and Apple, Samuel and Kharawala, Amrin and Duarte, Gustavo and Liaqat, Wasla and Fu, Yiwen and Li, Weijia and Chen, Yiyun and others},
  journal={Diagnostics},
  volume={12},
  number={3},
  pages={689},
  year={2022},
  publisher={MDPI}
}

@article{olaisen2023automatic,
  title={Automatic measurements of left ventricular volumes and ejection fraction by artificial intelligence: clinical validation in real time and large databases},
  author={Olaisen, Sindre and Smistad, Erik and Espeland, Torvald and Hu, Jieyu and Pasdeloup, David and {\O}stvik, Andreas and Aakhus, Svend and R{\"o}sner, Assami and Malm, Siri and Stylidis, Michael and others},
  journal={European Heart Journal-Cardiovascular Imaging},
  pages={jead280},
  year={2023},
  publisher={Oxford University Press US}
}

@article{elvas2023ai,
  title={AI-Based Aortic Stenosis Classification in MRI Scans},
  author={Elvas, Lu{\'\i}s B and {\'A}guas, Pedro and Ferreira, Joao C and Oliveira, Jo{\~a}o Pedro and Dias, Miguel Sales and Ros{\'a}rio, Lu{\'\i}s Br{\'a}s},
  journal={Electronics},
  volume={12},
  number={23},
  pages={4835},
  year={2023},
  publisher={MDPI}
}

@article{stonko2023review,
  title={A Review of Mature Machine Learning and Artificial Intelligence Enabled Applications in Aortic Surgery},
  author={Stonko, David P and Morrison, Jonathan J and Hicks, Caitlin W},
  journal={JVS-Vascular Insights},
  pages={100016},
  year={2023},
  publisher={Elsevier}
}

@article{doi:10.1016/j.jacep.2018.11.004,
author = {Gherardo Finocchiaro  and Michael Papadakis  and Gaia Tanzarella  and Harshil Dhutia  and Chris Miles  and Maite Tome  and Elijah R. Behr  and Sanjay Sharma  and Mary N. Sheppard },
title = {Sudden Death Can Be the First Manifestation of Hypertrophic Cardiomyopathy},
journal = {JACC: Clinical Electrophysiology},
volume = {5},
number = {2},
pages = {252-254},
year = {2019},
doi = {10.1016/j.jacep.2018.11.004},

URL = {https://www.jacc.org/doi/abs/10.1016/j.jacep.2018.11.004},
eprint = {https://www.jacc.org/doi/pdf/10.1016/j.jacep.2018.11.004}
}
%\bibliography{refsg,imageref}
%\bibliography{/Users/yudengdeng/Dropbox/Research/Refs/refsg} %This works for windows

\end{document}